\documentclass{aa}  

\usepackage{graphicx}
\usepackage{txfonts}
\begin{document}

   \title{JOYS: JWST Observations of Young protoStars}

   \subtitle{Outflows and accretion in the high-mass star-forming region IRAS\,23385+6053}

 \author{H.~Beuther
          \inst{1}
          \and
          E.~F.~van Dishoeck
          \inst{2}
          \and
          L.~Tychoniec
          \inst{3}
          \and
          C.~Gieser
          \inst{4}
          \and
          P.J.~Kavanagh
          \inst{5}
          \and
          G.~Perotti
          \inst{1}
          \and
          M.L.~van Gelder
          \inst{2}
           \and
          P.~Klaassen
          \inst{6}
           \and
          A.~Caratti o Garatti
          \inst{7}
           \and
          L.~Francis
          \inst{2}
           \and
          W.~R.~M.~Rocha
          \inst{2,8}
           \and
          K.~Slavicinska
          \inst{2,8}
           \and
          T.~Ray
          \inst{5}
           \and
          K.~Justtanont
          \inst{9}
          \and
          H.~Linnartz
          \inst{8}
          \and
          C.~Waelkens
          \inst{10}
          \and
          L.~Colina
          \inst{11}
          \and
          T.~Greve
          \inst{12}
          \and
          M.~G\"udel
          \inst{13,1,14}
          \and
          T.~Henning
          \inst{1}
          \and
          P.-O.~Lagage
          \inst{15}
          \and
          B.~Vandenbussche
          \inst{10}
          \and
          G.~\"Ostlin
          \inst{16}
          \and
          G.~Wright
          \inst{6}
}
   \institute{$^1$ Max Planck Institute for Astronomy, K\"onigstuhl 17,
     69117 Heidelberg, Germany, \email{beuther@mpia.de}\\
     $^2$ Leiden Observatory, Leiden University, P.O. Box 9513, NL 2300 RA Leiden, The Netherlands\\
     $^3$ European Southern Observatory, Karl-Schwarzschild-Strasse 2, 85748 Garching bei M\"unchen, Germany \\
     $^4$ Max-Planck-Institut for Extraterrestrial Physics, Gie\ss enbachstrasse 1, 85748 Garching\\
     $^5$ School of Cosmic Physics, Dublin Institute for Advanced Studies, 31 Fitzwilliam Place, Dublin 2, Ireland\\
     $^6$ UK Astronomy Technology Centre, Royal Observatory Edinburgh, Blackford Hill, Edinburgh EH9 3HJ, UK\\
     $^7$ INAF-Osservatorio Astronomico di Capodimonte, Salita Moiariello 16, I-80131 Napoli, Italy\\
     $^8$ Laboratory for Astrophysics, Leiden Observatory, Leiden University, P.O. Box 9513, NL 2300 RA Leiden, The Netherlands\\
     $^9$ Chalmers University of Technology, Onsala Space Observatory, 439 92 Onsala, Sweden\\
     $^{10}$ Instituut voor Sterrenkunde, KU Leuven, Celestijnenlaan 200D, Bus-2410, 3000 Leuven, Belgium\\
     $^{11}$ Centro de Astrobiologıa (CAB, CSIC-INTA), Carretera de Ajalvir, E-28850 Torrejon de Ardoz, Madrid, Spain\\
     $^{12}$ DTU Space, Technical University of Denmark. Building 328, Elektrovej, 2800 Kgs.~Lyngby, Denmark\\
     $^{13}$ Dept. of Astrophysics, University of Vienna, T\"urkenschanzstr. 17, A-1180 Vienna, Austria\\
     $^{14}$ ETH Z\"urich, Institute for Particle Physics and Astrophysics, Wolfgang-Pauli-Str. 27, 8093 Z\"urich, Switzerland\\
     $^{15}$ Universite Paris-Saclay, Universite de Paris, CEA, CNRS, AIM, F-91191 Gif-sur-Yvette, France\\
     $^{16}$ Department of Astronomy, Oskar Klein Centre; Stockholm University; SE-106 91 Stockholm, Sweden
   }

   \date{Version of \today}

 
  \abstract
   {Understanding the earliest stages of star formation, and setting that into context with the general cycle of matter in the interstellar medium, is a central aspect of research with the James Webb Space Telescope (JWST).}
   {The JWST program JOYS (JWST Observations of Young protoStars) aims at characterizing the physical and chemical properties of young high- and low-mass star-forming regions, in particular the unique mid-infrared diagnostics of the warmer gas and solid-state components. We present early results from the high-mass star formation region IRAS\,23385+6053.}
   {The JOYS program uses the Mid-Infrared Instrument (MIRI) Medium Resolution Spectrometer (MRS) with its Integral Field Unit (IFU) to investigate a sample of high- and low-mass star-forming protostellar systems.}
   {The full 5 to 28\,$\mu$m MIRI MRS spectrum of IRAS\,23385+6053 shows a plethora of interesting features. While the general spectrum is typical for an embedded protostar, we see many atomic and molecular gas lines boosted by the higher spectral resolution and sensitivity compared to previous space missions. Furthermore, ice and dust absorption features are also present. Here, we focus on the continuum emission, outflow tracers like the H$_2$(0--0)S(7), [FeII]($^4F_{9/2}$--$^6D_{9/2}$) and [NeII]($^2P_{1/2}-^2P_{3/2}$) lines as well as the potential accretion tracer Humphreys $\alpha$ H{\sc{i}}(7--6). The short-wavelength MIRI data resolve two continuum sources A and B, where mid-infrared source A is associated with the main mm continuum peak. The combination of mid-infrared and mm data reveals a young cluster in its making. Combining the mid-infrared outflow tracer H$_2$, [FeII] and [NeII] with mm SiO data shows a complex interplay of at least three molecular outflows driven by protostars in the forming cluster. Furthermore, the Humphreys $\alpha$ line is detected at a 3--4$\sigma$ level towards the mid-infrared sources A and B. Following \citet{rigliaco2015}, one can roughly estimate both accretion luminosities and corresponding accretion rates between $\sim$2.6$\times 10^{-6}$ and $\sim$0.9$\times 10^{-4}$\,$M_{\odot}yr^{-1}$. This is discussed in the context of the observed outflow rates.} 
   {The analysis of the MIRI MRS observations for this young high-mass star-forming region reveals connected outflow and accretion signatures. Furthermore, they outline the enormous potential of JWST to boost our understanding of the physical and chemical processes during star formation.}


   \maketitle
%

\section{Introduction}
\label{intro}

\begin{figure*}[htb]
\includegraphics[width=0.99\textwidth]{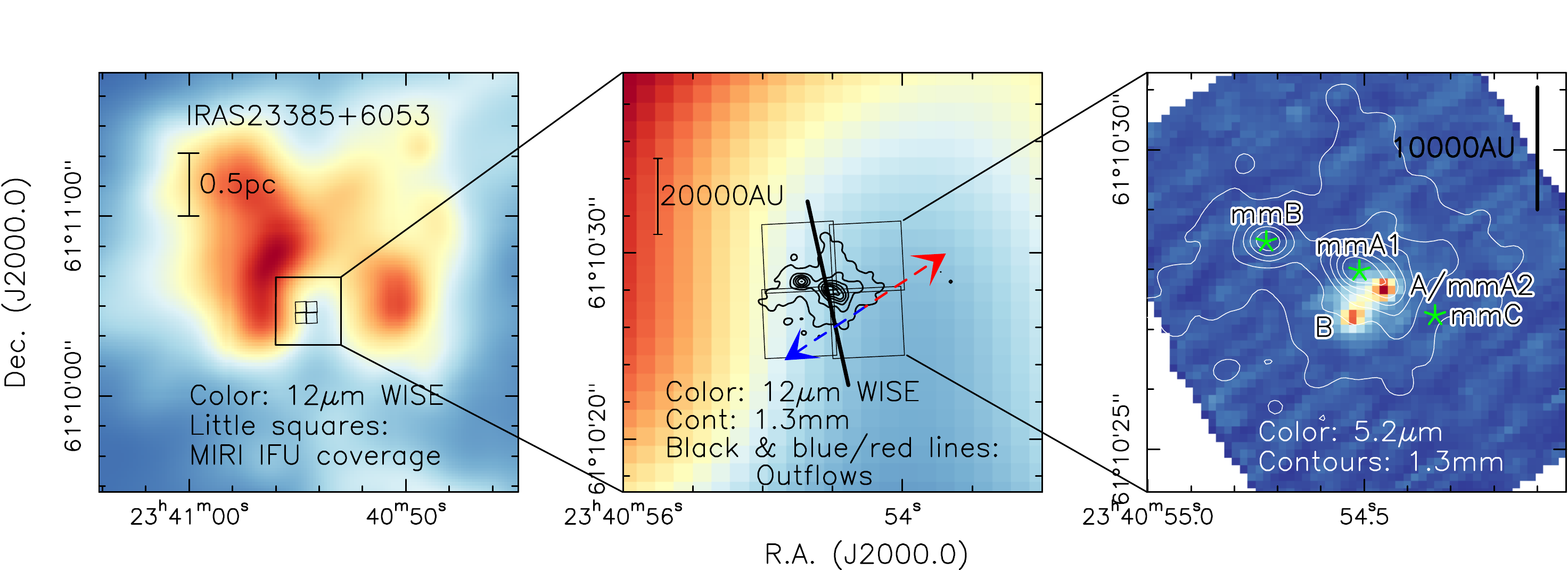}
\caption{Overview of the region around IRAS\,23385+6053. The left panel shows a large-scale overview while the middle and right panels present  zoom-ins. The color-scale in the left and middle panels show the 12\,$\mu$m emission from WISE at an angular resolution of $6.5''$ \citep{wright2010}. The little squares outline the field of view of the MIRI IFU mosaic in channel 1 (each square $3.7''\times 3.7''$). The color-scale in the right panel shows the new MIRI 5.2\,$\mu$m continuum emission (see also Fig.~\ref{continuum}). The contours in the middle and right panels present the 1.3\,mm continuum emission \citep{beuther2018b,cesaroni2019}. The contours start at the 5$\sigma$ level of 0.55\,mJy\,beam$^{-1}$ and continue in 25$\sigma$ steps. The black and red/blue lines in the middle panel show the directions of outflows identified in \citet{molinari1998b} and \citet{cesaroni2019}, respectively. The right panel marks the mid-infrared sources A and B as well as the mm sources labeled "mm" with the corresponding source letters from \citet{cesaroni2019}. Mid-infrared source A and mm source mmA2 are spatially co-located. A linear scale-bar is shown in all panels.}
\label{overview} 
\end{figure*} 

The earliest phases of protostellar evolution are deeply embedded in their natal cloud cores. Many important physical and chemical processes occur there, including the infall of the envelope, the formation of disks and outflows, the main accretion phase and growth of the final star, as well as the chemical enrichment of the disk directly impacting planet formation. Because of the very high extinction $A_V$ of several hundreds to even thousands of magnitudes, these deeply embedded phases have to be studied at (mid-)infrared and (sub)mm wavelengths. While (sub)mm interferometers have allowed detailed imaging of the cold material in such star-forming cores (for a review see \citealt{motte2018}), mid-infrared continuum and line imaging of the warmer material at sub-arcsecond spatial resolution has been barely feasible so far. Only a few space-based studies at lower angular resolution with the Infrared Space Observatory (ISO) and Spitzer are reported (e.g., \citealt{whittet1996,vandishoeck1998b,gibb2000,molinari2008b,an2011}). The most comprehensive ISO legacy study on high-mass protostars is presented in \citet{gibb2004c}.

The situation has now changed dramatically with the advent of the James Webb Space Telescope (JWST), launched Dec.~25, 2021, just little more than a year ago. Scientific data delivery began only in July 2022, and we present the first results from the European MIRI (Mid-Infrared Instrument) Guaranteed Time Program JOYS (JWST Observations of Young protoStars, PI: E.~F.~van Dishoeck, pid 1290).

The JOYS program observes a total of about two dozen low- to high-mass star-forming regions with the MIRI Medium Resolution Spectrometer (MRS) and its Integral Field Unit (IFU) to address a broad range of scientific topics. Evolutionary stages from very young embedded protostars within infrared dark clouds to Class 0/I  protostars as well as high-mass protostellar objects and hot molecular cores are being covered by the JOYS project. The spatial resolution of the MIRI instrument of $\sim 0.19''$, $\sim 0.58''$ and $\sim 0.96''$ at 5, 15 and 25\,$\mu$m, respectively, corresponds, at typical distances of the low-mass cores of $\sim$150\, pc, to linear resolution elements of $\sim$29, $\sim$87 and $\sim$144\,au, respectively. For high-mass regions at more typical distances of $\sim$3\,kpc, the linear resolution corresponds roughly to 570, 1740 and 2880\,au.  

\begin{figure}[htb]
\includegraphics[width=0.49\textwidth]{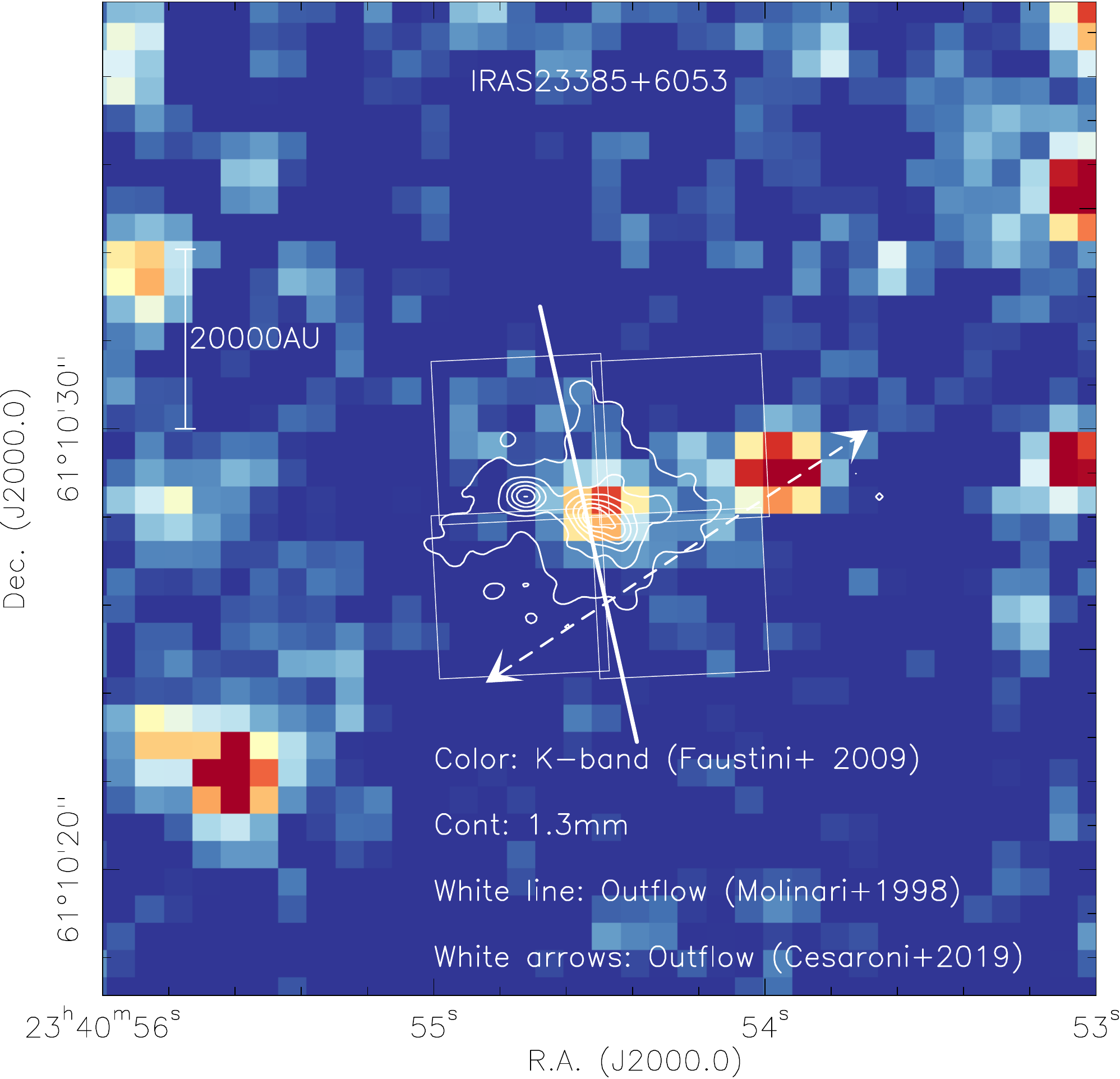}
\caption{Infrared K-band data are shown in color (taken with the Palomar 60-inch telescope and the NICMOS-3 array, \citealt{faustini2009}). The contours outline the 1.3\,mm emission \citep{beuther2018b,cesaroni2019}. Contour levels start at 5$\sigma$ (0.55\,mJy\,beam$^{-1}$) and continue in 25$\sigma$ steps. The white full and dashed lines show the directions of outflows identified in \citet{molinari1998b} and \citet{cesaroni2019}, respectively.}
\label{23385_k} 
\end{figure} 


We are using a set of key diagnostics in the mid-infrared band between 5 and 28\,$\mu$m such as the continuum spectral energy distribution (SED), a range of atomic and molecular gas lines as well as several solid state bands. More details about the JOYS project will be presented in a forthcoming publication by van Dishoeck et al. (in prep.). 

\begin{figure*}[htb]
\includegraphics[width=0.99\textwidth]{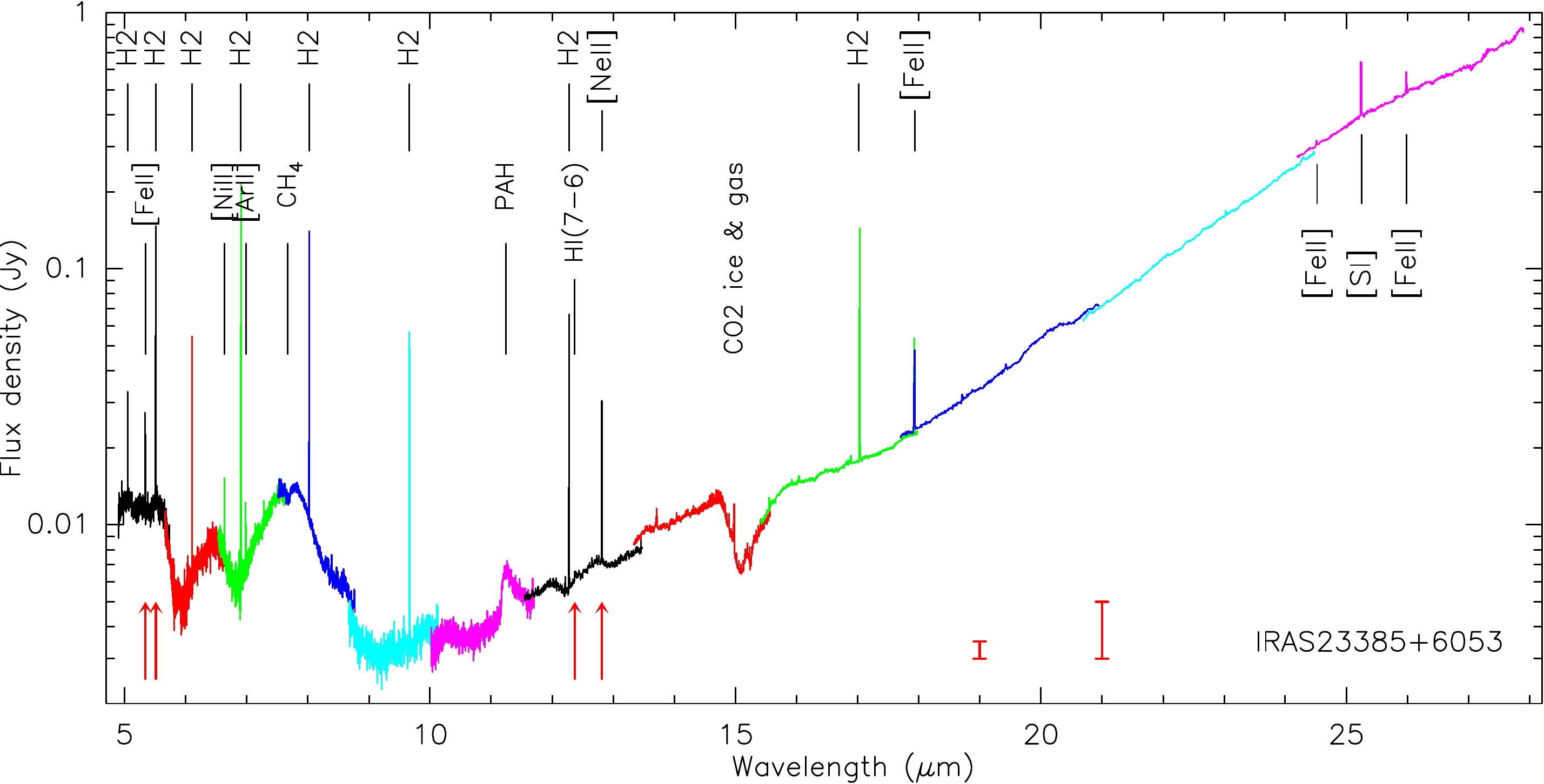}
\caption{Full MIRI spectrum extracted towards the combined sources A and B (central position: R.A.(J2000.0) 23:40:54.49, Decl.~61:10:27.4) within a diameter of $2.4''$). The color coding separates the different sub-bands from channel 1 short to channel 4 long. Prominent strong atomic and molecular lines are labeled (see Table \ref{features}). The four lines mainly discussed in this paper are highlighted by the red arrows at the bottom. The data are not background-subtracted. The errorbars at 19 and 21\,$\mu$m show the approximate $1\sigma$ rms below and above 20\,$\mu$m of $\sim$0.5 and $\sim$2\,mJy (note the log-scale), respectively (Sect.~\ref{observations}).}
\label{spectrum} 
\end{figure*} 

Here, we focus on analysing and interpreting JWST data of the first source fully observed within the JOYS program, the high-mass protostellar object IRAS\,23385+6053. This high-mass star-forming region was originally identified by IRAS color selection and association with H$_2$O maser emission \citep{casoli1986,cesaroni1988,wouterloot1989,palla1991}. Early interferometric investigations revealed that the region is embedded in a comparatively dark cloud core surrounded by larger-scale mid-infrared emission \citep{molinari1998b}. Figure \ref{overview} presents an overview of the region at 12\,$\mu$m (left panel, from the Wide-field Infrared Survey Explorer -- WISE -- mission at $6.5''$ resolution, \citealt{wright2010}) and 1.3\,mm emission (right panel, \citealt{beuther2018b}, \citealt{cesaroni2019}) showing the large-scale extended mid-infrared emission that weakens at the positions of the central bright mm emission. The outline of the 4-field MIRI-IFU coverage is shown there as well. The $v_{\rm lsr}$ of the region is $-50.2$\,km\,s$^{-1}$. At a kinematic distance of 4.9\,kpc, the total luminosity measured with the IRAS mission accounts for $1.6\times 10^4$\,$L_{\odot}$ \citep{molinari1998b}. However, higher-spatial-resolution Spitzer/MIPSGAL observations revealed that only $\sim$3$\times 10^3$\,$L_{\odot}$ are associated with the central core, consistent with a high-mass protostar in the main accretion phase \citep{molinari2008b}. Infrared imaging by \citet{faustini2009} exhibits an infrared cluster in the surroundings of the region. A weak K-band source is detected in their data toward the very center (Fig.~\ref{23385_k}). An H$_2$ emission-line study at 2.12\,$\mu$m revealed shocked H$_2$ emission in the immediate surroundings of the central mm-core \citep{wolf-chase2012}. CH$_3$OH maser emission at 44 and 95\,GHz confirm shock emission in IRAS\,23385+6053 \citep{kurtz2004,wolf-chase2012}. The most recent high-spatial resolution mm observations ($\sim$0.4$''$) were conducted within the IRAM NOEMA (Northern Extended Millimeter Array) large program CORE that studied the physical and chemical properties of 20 high-mass star-forming regions \citep{beuther2018b,gieser2021}. A detailed case study of IRAS\,23385+6053 from the CORE project is presented in \citet{cesaroni2019}. Based on the line and continuum emission, \citet{cesaroni2019} identified six cores, a new outflow candidate in the northwest-southeast direction (Fig.~\ref{overview}, middle panel) and a tentative detection of a disk around a $\sim$9\,$M_{\odot}$ protostar within the most massive mm core mmA1 (Fig.~\ref{overview}, right panel). 

This first paper in the JOYS-series presents an overview of the overall MIRI spectrum and then focuses on the continuum emission and a few selected molecular and atomic lines (H$_2$(0--0)S(7), [FeII]($^4F_{9/2}$--$^6D_{9/2}$) and Humphreys $\alpha$ H{\sc{i}}(7--6)). The data outline the general source structure, and we investigate the outflow properties and potential accretion signatures. Complementary studies will discuss the overall extended emission of the diverse set of detected atomic and H$_2$ lines and set them in relation to the molecular gas properties derived at mm wavelengths (Gieser et al.~in prep.), the molecular emission around the central disk candidate (Francis et al.~in prep.) and the ice properties in the region (Rocha et al.~in prep.).

\section{Observations}
\label{observations}

IRAS\,23385+6053 was observed within the European MIRI GTO program JOYS (proposal id.~1290) for 2.12\,h on August 22, 2022. The entire MIRI bandpass from $\sim$4.9 to $\sim$28.0\,$\mu$m was covered in a 4-field mosaic (Fig.~\ref{overview}) around the central position of R.A.~(J2000.0) 23:40:54.497 and Dec.~(J2000.0) +61:10:27.83. The total exposure time in each the three MIRI spectral filters was 200\,sec, and parallel off-source imaging was conducted in the F1500W filter. A complementary background field was observed for overall background subtraction that is needed, e.g., an analysis of the broad-band ice features. While the full-wavelength spectrum in Fig.~\ref{spectrum} shows the non-background subtracted data, all individual narrow gas spectral line data presented in this paper were continuum subtracted by fitting a polynomial baseline to narrow wavelength ranges around these lines.

\begin{figure*}[htb]
\includegraphics[width=0.99\textwidth]{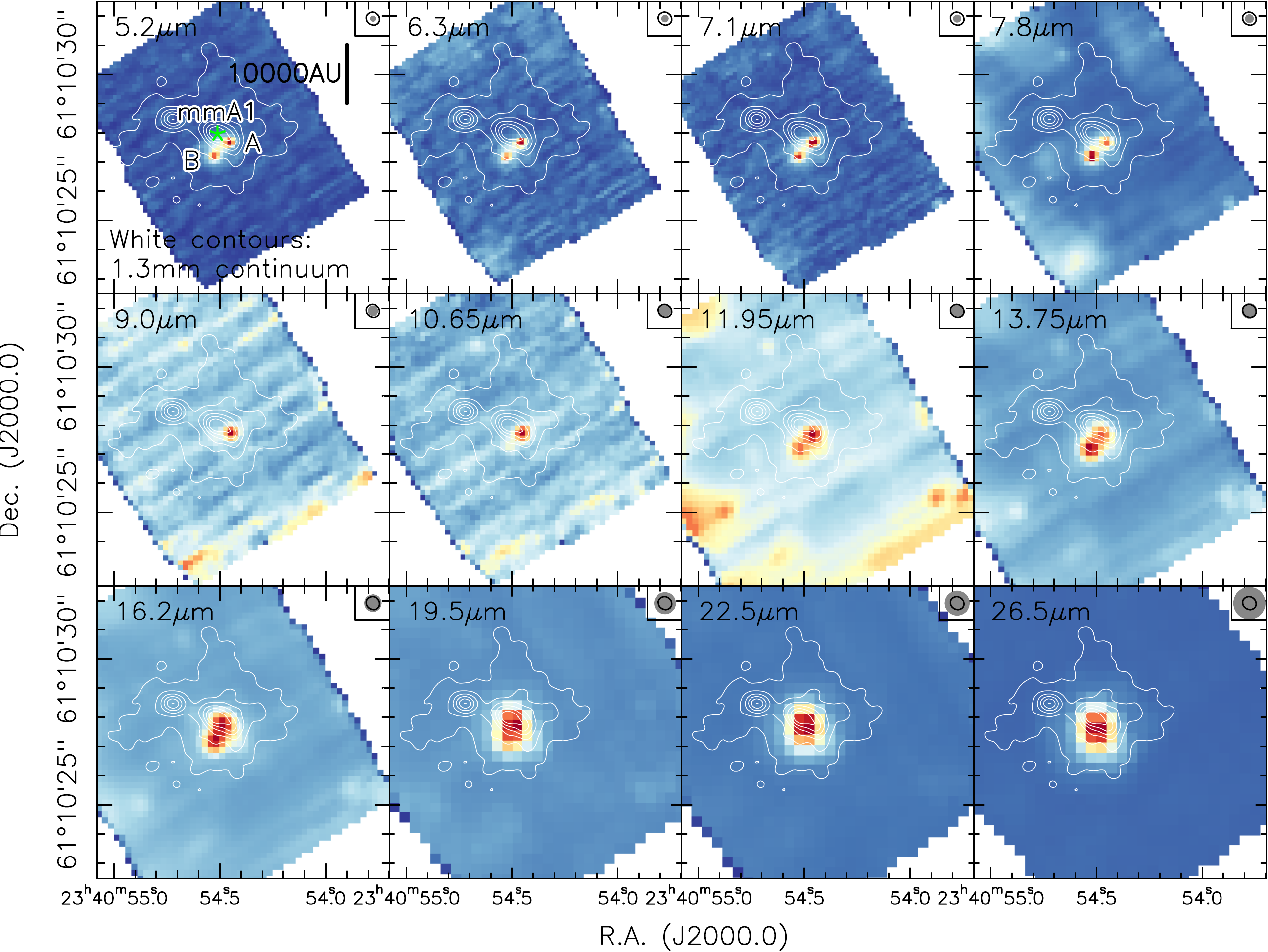}
\caption{Continuum images of IRAS\,23385+6053. The color-scale presents the continuum emission extracted from the MIRI IFU data cube in a line-free part of each spectral setting centered on the wavelength labeled in each panel. The contours outline the 1.3\,mm emission \citep{beuther2018b,cesaroni2019} starting at the 5$\sigma$ level of 0.55\,mJy\,beam$^{-1}$ and continuing in 25$\sigma$ steps. In the top-right of each panel, the corresponding spatial resolution is shown (grey: mid-infrared, line: 1.3\,mm). The top-left panel labels the two main mid-infrared sources A and B as well as marks the main mm peak mmA1 from \citet{cesaroni2019} with a green star. A linear scale-bar is shown as well. Note that mid-infrared source A and B are well separated up to 8\,$\mu$m, but not at longer wavelengths.}
\label{continuum} 
\end{figure*} 

We reduced the data using a development version of the JWST calibration pipeline \citep{bushouse2022} with v.~11.16.16 and CRDS (Calibration Reference Data System) context 'jwst\_1017.pmap', respectively. We used this development version as it includes cosmic ray shower flagging in the \texttt{jump} step of the \texttt{Detector1Pipeline}.

The pipeline includes fringe correction via \texttt{Spec2Pipeline} using a fringe flat field derived from spatially extended sources (Mueller et al.~in prep). We further reduced the fringing by applying the \texttt{residual\_fringe} step (Kavanagh et al.~in prep.), which is included in the JWST calibration pipeline package but not switched on by default.

As discussed in \citet{pontoppidan2022}, the accuracy of spacecraft pointing information for MRS exposures can be affected by guide star catalog errors and roll uncertainty. We corrected our MRS exposures following the same procedure as in that paper, i.e., using identified Gaia-DR3\footnote{\url{https://www.cosmos.esa.int/web/gaia/dr3}} sources in the MIRI simultaneous imaging field, leading to a pointing adjustment of $1.6077''$ in Right Ascension and $0.3485''$ in Declination. This correction is still subject to uncertainties in the relative astrometry between the imager and the MRS, which is expected to be accurate to $\leq 0.1''$ (Patapis et al.~in prep.).

Finally, we created spectral cubes for each of the 12 MRS sub-bands (four MIRI channels with each containing a "short", "medium" and "long" grating setting) using the level 2 data products with corrected astrometry from which spectra were extracted. We applied the post-pipeline residual fringe correction tool to all spectra to account for high frequency fringe residuals, thought to originate in the MRS dichroics, which are not effectively removed by the \texttt{residual\_fringe} step in the pipeline (Kavanagh et al.~in prep). The $1\sigma$ rms of the spectrum presented in Figure \ref{spectrum} is typically around 0.5\,mJy below 20\,$\mu$m and rises to a few mJy at longer wavelengths.

\begin{figure*}[htb]
\includegraphics[width=0.99\textwidth]{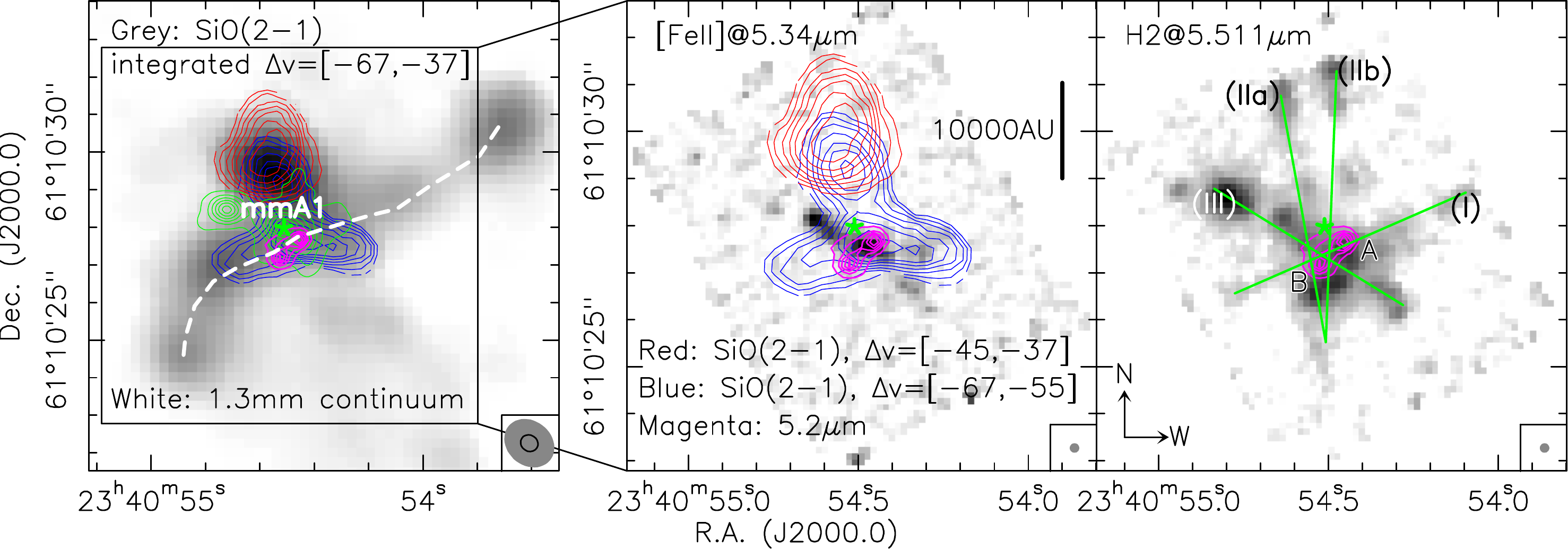}
\caption{Outflow images in NOEMA SiO(2-1) (integrated emission, left panel), JWST [FeII]($^4F_{9/2}$--$^6D_{9/2}$)@5.34\,$\mu$m (middle panel) and JWST H$_2$(0--0)S(7)@5.511\,$\mu$m (right panel). The red and blue contours show the red- and blue-shifted NOEMA SiO(2--1) emission in the labeled velocity regimes ($v_{\rm lsr}=-50.2$\,km\,s$^{-1}$). Contour levels are from 30 to 90\% (step 10\%) of the respective integrated peak emission. The magenta contours show the 5.2\,$\mu$m continuum emission (steps 20 to 90\% by 15\%). The green contours in the left panel outline the 1.3\,mm continuum starting at the 20$\sigma$ levels (1$\sigma$=1.13\,mJy\,beam$^{-1}$). A linear scale bar is presented in the middle panel, and the green lines in the right panel outline and label potential outflows. The two mid-infrared sources are labeled in the right panel as well. The dashed line in the left panel marks a potentially precessing outflow corresponding to outflow (I). A green star in all panels marks the main mm peak position labeled mmA1 in \citet{cesaroni2019}. North and west are labeled in the right panel. The resolution elements are shown at the bottom right of each panel, left SiO(2--1) in grey and 1.3\,mm continuum as line, middle and right [FeII] and H$_2$, respectively.}
\label{outflows} 
\end{figure*} 

Complementary 1.3\,mm continuum data are used from the CORE project \citep{beuther2018b,cesaroni2019}. The angular resolution and sensitivity of these data are $0.48''\times 0.43''$ and 0.11\,mJy\,beam$^{-1}$, respectively. Furthermore, SiO(2--1) observations are taken from the CORE follow-up project CORE+ (PI: Caroline Gieser). This project aims to study the shock and deuterium chemistry in the CORE sample. IRAS\,23385+6053 is part of the CORE+ pilot study (Gieser et al.~in prep.). The corresponding angular resolution and sensitivity of the SiO(2--1) data are $1.4'' \times 1.2''$ and $\sim$0.12\,K per 0.8\,km\,s$^{-1}$, respectively.

\section{Results}

\subsection{General spectral features}

Figure \ref{spectrum} presents the spectrum extracted towards the combined emission from the two mid-infrared sources A and B (Sect.~\ref{sec_continuum} and Fig.~\ref{continuum}). We present the combined spectrum from sources A and B because at wavelengths longward of $\sim$13\,$\mu$m, they cannot be spatially separated anymore (Fig.~\ref{continuum}). The color-coding corresponds to the four MIRI channels with the corresponding three grating settings each, hence 12 individual sub-bands. It is evident that JWST can now take high-quality mid-infrared spectra of sources as weak as 10\,mJy in short integration times, more than a factor 1000 fainter than previous data. Several features can directly be identified (see Table \ref{features}). Between $\sim$5 and $\sim$11.5\,$\mu$m, the spectrum is dominated by well-known structures that stem from several ice and solid-state features (e.g., \citealt{gibb2004c,yang2022,mcclure2023}). In particular, the silicate band around 10\,$\mu$m shows strong absorption, almost reaching full saturation. Going to longer wavelengths, the general emission rises as expected from warm dust emission. On top of the continuum emission, ice and solid-state features as well as many strong emission lines can be identified, in particular from H$_2$ and atomic lines from [FeII], [ArII], [NiII], [NeII] and [SI] (all labeled in Fig.~\ref{spectrum}). In addition to these strong lines, also a few weaker molecular lines can be identified, in particular those of CO$_2$, C$_2$H$_2$, CH$_4$ and HCN (Francis et al.~in prep.). While the H$_2$ and atomic line emission is extended, the molecular emission mainly stems from the environment of the continuum sources. Selected H$_2$ and [FeII] line images are presented in Sect.~\ref{lines}, whereas the remaining gas and ice analysis will be discussed in accompanying papers (Sect.~\ref{intro}). 

\begin{table}[htb]
\caption{Main spectral features.}
\begin{tabular}{lr|lr}
  \hline \hline
Feature & $\lambda$ & Feature & $\lambda$ \\
        & ($\mu$m) & & ($\mu$m) \\
\hline
H$_2$ S(8) & 5.053 & PAH & 11.238 \\

[FeII] & 5.340 & H$_2$ S(2) & 12.279 \\
H$_2$ S(7) & 5.511 & H{\sc i} & 12.370 \\
H$_2$ S(6) & 6.109 & [NeII] & 12.814 \\

[NiII] & 6.636 & CO$_2$ ice/gas & 14.5--15.5 \\
H$_2$ S(5) & 6.910 & H$_2$ S(1) & 17.035 \\

[ArII] & 6.985 & [FeII] & 17.936 \\
CH$_4$ & 7.670 & [FeII] & 24.519 \\
H$_2$ S(4) & 8.025  & [SI] & 25.249 \\
H$_2$ S(3) & 9.665 & [FeII] & 25.988 \\
\hline \hline
\end{tabular}
\label{features}
\end{table}

\begin{figure*}[htb]
\includegraphics[width=0.99\textwidth]{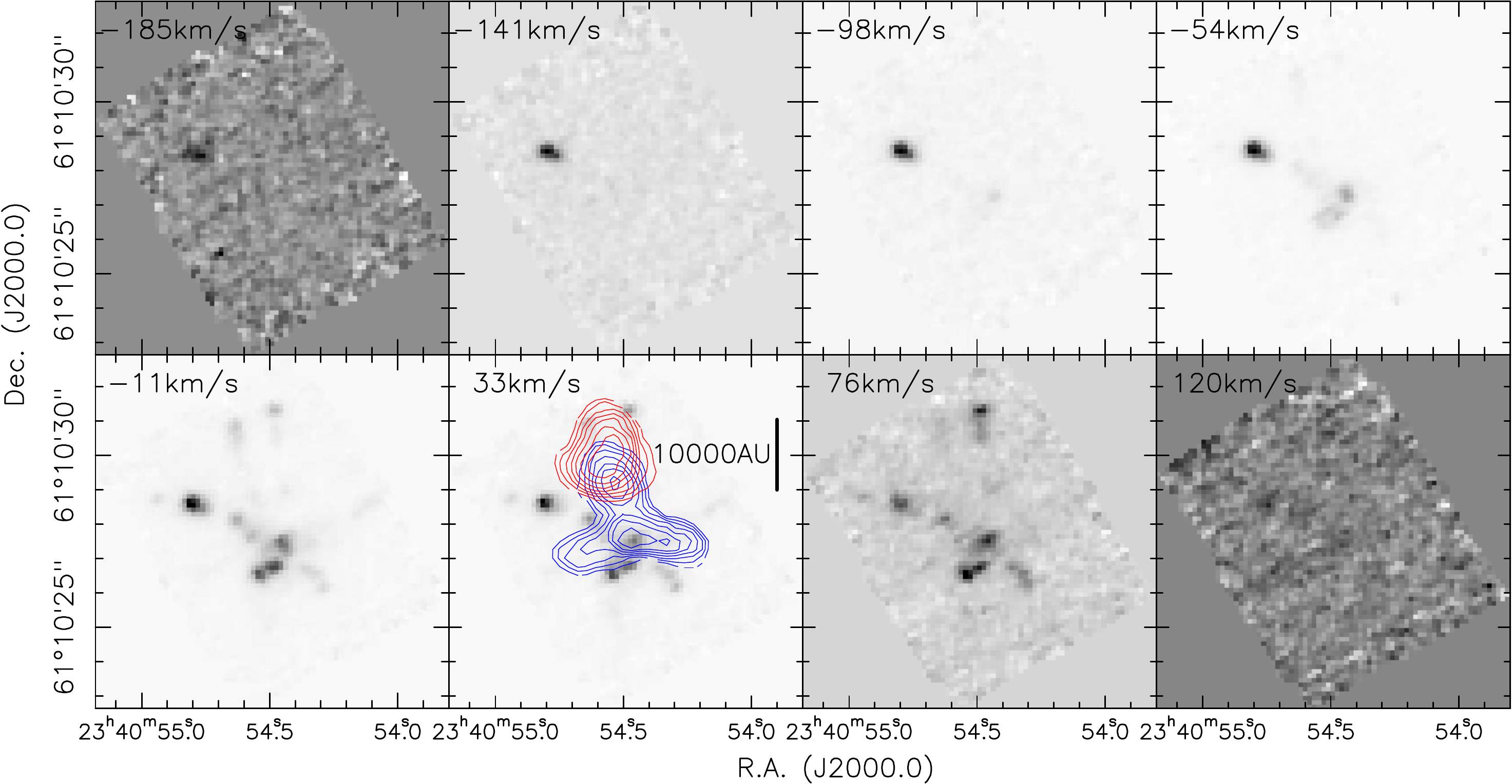}
\caption{Channel map of the H$_2$(0--0)\,S(7) line at 5.511\,$\mu$m. The velocity of each channel is marked in units of km\,s$^{-1}$ in the top-left corner of the panels. The red and blue contours in the 33\,km\,s$^{-1}$ channel correspond to the same red- and blue-shifted SiO(2--1) emission as presented in Fig.~\ref{outflows}. A linear scale-bar is shown in that panel as well. The $v_{\rm lsr}$ is $-50.2$\,km\,s$^{-1}$.}
\label{channel_h2} 
\end{figure*} 

\subsection{Continuum emission}
\label{sec_continuum}

Figure \ref{continuum} presents the mid-infrared continuum emission extracted for each of the twelve MIRI sub-bands averaged over wavelength ranges without significant line emission. The central wavelengths of each image are marked in all panels. We show the mid-infrared emission in comparison to the mm continuum emission from the CORE project (\citealt{beuther2018b}, \citealt{cesaroni2019}, see also Fig.~\ref{overview} for more detailed source labels). At the short wavelengths, one clearly identifies two mid-infrared sources associated with the main mm continuum core. That is also the area where the K-band emission is detected (\citealt{faustini2009},  Fig.~\ref{23385_k}). The north-western mid-infrared source A is closer to the main mm peak position (mmA1 in \citealt{cesaroni2019}) whereas the south-eastern source B is offset from mid-infrared source A by $\sim$0.67$''$. At the given distance of 4.9\,kpc, the angular offset between the two mid-infrared sources A and B corresponds to a projected linear separation of $\sim$3280\,au.

While the offset between the mid-infrared sources A and B is well determined, the absolute positional offsets between the mm and mid-infrared data is less well constrained. Positions from mm interferometer observations like the NOEMA CORE data \citep{beuther2018b} are largely constrained by the phase error, and should be better than $0.1''$ for the presented mm data. The corresponding 3\,mm continuum data do also agree very well with the 1.3\,mm data presented here (Gieser et al.~in prep.). In contrast to that, the mid-infrared positional accuracy depends on the absolute JWST pointing, and the accuracy with which the data can be improved by using known GAIA positions during the data reduction. As outlined in Sect.~\ref{observations}, after correlating the parallel imaging data with GAIA observations, the data were shifted in R.A./Dec.~by $\sim$1.6077$''/0.3485''$. In addition to the alignment with Gaia, this correction needs additional knowledge of the alignment of the MIRI imager with the MIRI IFU. This alignment should be very good and accurate to $0.1''$ (Sect.~\ref{observations}). This implies that mid-infrared source A should indeed be very close to the elongation of the main mm peak (the mm elongation is labeled peak mmA2 in \citealt{cesaroni2019}) whereas the offset of mid-infrared source B to the south-east should be real as well.

Going to longer wavelengths, in the spectral range of the silicate feature between roughly 9 and 11\,$\mu$m, source B almost disappears. This indicates that the silicate absorption feature toward source B is even deeper and entirely absorbs the continuum emission, whereas for source A some continuum emission still remains. A detailed analysis of all solid state features will be presented in Rocha et al.~(in prep.). Moving to even longer wavelengths beyond the silicate feature, source B re-appears. From $\sim$16\,$\mu$m onward, the further decreasing spatial resolution does not allow us to separate the two mid-infrared sources anymore, and they merge into a single source at the long-wavelength end of the MIRI bandpass.

\subsection{H$_2$, [FeII] and [NeII] line emission}
\label{lines}

The MIRI spectral line data now allow us to get a better understanding of the underlying outflow structures in IRAS\,23385+6053. While \citet{molinari1998b} reported one outflow roughly in the north-south direction, \citet{cesaroni2019} recently suggested that a second outflow in the northwest-southeast direction should exist (Fig.~\ref{overview} right panel). The shocked H$_2$ emission as well as CH$_3$OH maser emission at 44 and 95\,GHz additionally confirm the existence of shocked gas \citep{kurtz2004,wolf-chase2012}. Here, we focus on the high-spatial-resolution H$_2$(0--0)\,S(7) line at 5.511\,$\mu$m and the [FeII]($^4F_{9/2}$--$^6D_{9/2}$) line at 5.34\,$\mu$m. Furthermore, we compare the results with SiO(2--1) data at 3.6\,mm wavelengths obtained with NOEMA as part of the CORE+ project (Gieser et al.~in prep.). Figure \ref{outflows} presents a compilation of the three datasets. For the SiO(2--1) data, we show the integrated emission ($\pm 15$\,km\,s$^{-1}$ around the $v_{\rm lsr}=-50.2$\,km\,s$^{-1}$) as well as separated blue- and red-shifted high-velocity gas emission, whereas for [FeII] and H$_2$ we present the integrated emission. 

\begin{figure*}[htb]
\includegraphics[width=0.99\textwidth]{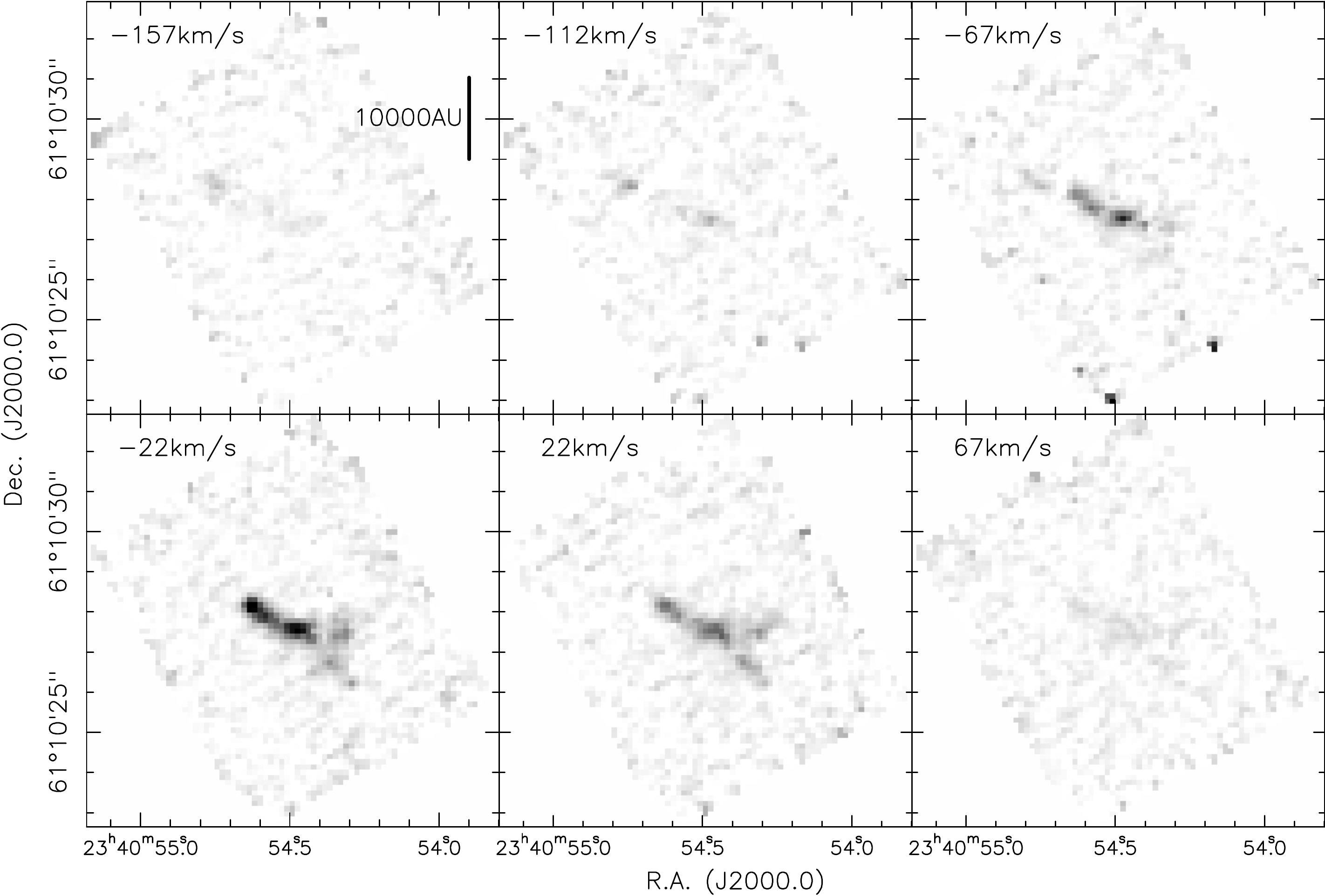}
\caption{Channel map of the Fe[II]($^4F_{9/2}$--$^6D_{9/2}$) line at 5.34\,$\mu$m. The velocity of each channel is marked units of km\,s$^{-1}$ in the top-left corner of the panels. A linear scale-bar is shown in the top-left panel. The $v_{\rm lsr}$ is $-50.2$\,km\,s$^{-1}$.}
\label{channel_fe} 
\end{figure*} 

Starting with the mm SiO emission, these data are compatible with at least two outflows. The integrated SiO(2--1) emission (grey-scale in Fig.~\ref{outflows} left) is consistent with the northwest-southeast outflow ((I) in Fig.~\ref{outflows} right panel) proposed by \citet{cesaroni2019}. The SiO emission also shows a more extended integral-shaped structure in the northwest-southeast direction (dashed line in Fig.~\ref{outflows} left, see discussion in Sect.~\ref{outflows_discussion}). In comparison,  the red- and blue-shifted SiO emission rather points at an outflow more aligned with the north-south direction ((IIa) \& (IIb) in Fig.~\ref{outflows} right panel) as suggested by \citet{molinari1998b}. However, especially the blue-shifted SiO emission shows extensions towards the east and west that may also be associated with other outflow structures. 

The [FeII] emission (Fig.~\ref{outflows} middle) exhibits a very different structure. It is dominated by a collimated jet-like outflow in the northeast-southwest direction ((III) in Fig.~\ref{outflows} right panel). But it also shows some extension toward the west, spatially associated with the blue-shifted SiO emission.

Last but not least, the H$_2$ data (Fig.~\ref{outflows} right) show emission structures associated with all the before mentioned outflows. There is strong emission associated with the northeast-southwest jet (III) traced also in [FeII], but it shows additional emission towards the northwest, similar to the integrated SiO emission (I). Furthermore, the H$_2$ data also exhibit strong emission towards the north (labeled (IIa) and (IIb) in Fig.~\ref{outflows} right panel), and slightly weaker in the south, that corresponds to the red- and blue-shifted SiO emission. 

In addition to the integrated emission, with a spectral resolving power of $\sim$3500 in channel 1A (\citealt{labiano2021}\footnote{https://jwst-docs.stsci.edu/jwst-mid-infrared-instrument/miri-observing-modes/miri-medium-resolution-spectroscopy}), we can resolve velocity elements down to $\sim$86\,km\,s$^{-1}$ (corrected for radial and heliocentric velocities). With an approximate Nyquist-sampling the channel separation is $\sim$44\,km\,s$^{-1}$. Figure \ref{channel_h2} presents a channel map for the H$_2$ line at 5.511\,$\mu$m. Interestingly, the gas velocities are indeed so high that we can velocity-resolve the different components. With the $v_{\rm{lsr}}$ of IRAS\,23385+6053 of $-50.2$\,km\,s$^{-1}$ \citep{beuther2018b}, we are covering a velocity range of roughly $\pm 140$\,km\,s$^{-1}$. For comparison, we also present the channel map of the [FeII] line at 5.34\,$\mu$m in Figure \ref{channel_fe}. Again, we can resolve blue- and red-shifted gas, however with a slightly smaller velocity spread of around $\pm 100$\,km\,s$^{-1}$.

Investigating the velocity structures in a bit more detail, we find that the H$_2$ emission features towards the north are all redshifted. This is consistent with the red-shifted SiO emission -- although SiO is at lower velocities -- that is also found in the north (Fig.~\ref{outflows}).  In contrast to that, the strong H$_2$ emission in the northeast is seen in all velocity channels, hence red- and blue-shifted. If we assign that feature to the northeast-southwest jet (III) that is also seen nicely in the [FeII] emission (Fig.~\ref{outflows} \& \ref{channel_fe}), this implies that this jet should be aligned close to the plane of the sky. The outflow (I) in the northwest-southeast direction can also be identified in the H$_2$ channel map (Fig.~\ref{channel_h2}), however, barely in blue-shifted gas and more easily at red-shifted velocities. This velocity structure is not well recovered in the SiO emission (Fig.~\ref{outflows}).

\begin{figure*}[htb]
\includegraphics[width=0.99\textwidth]{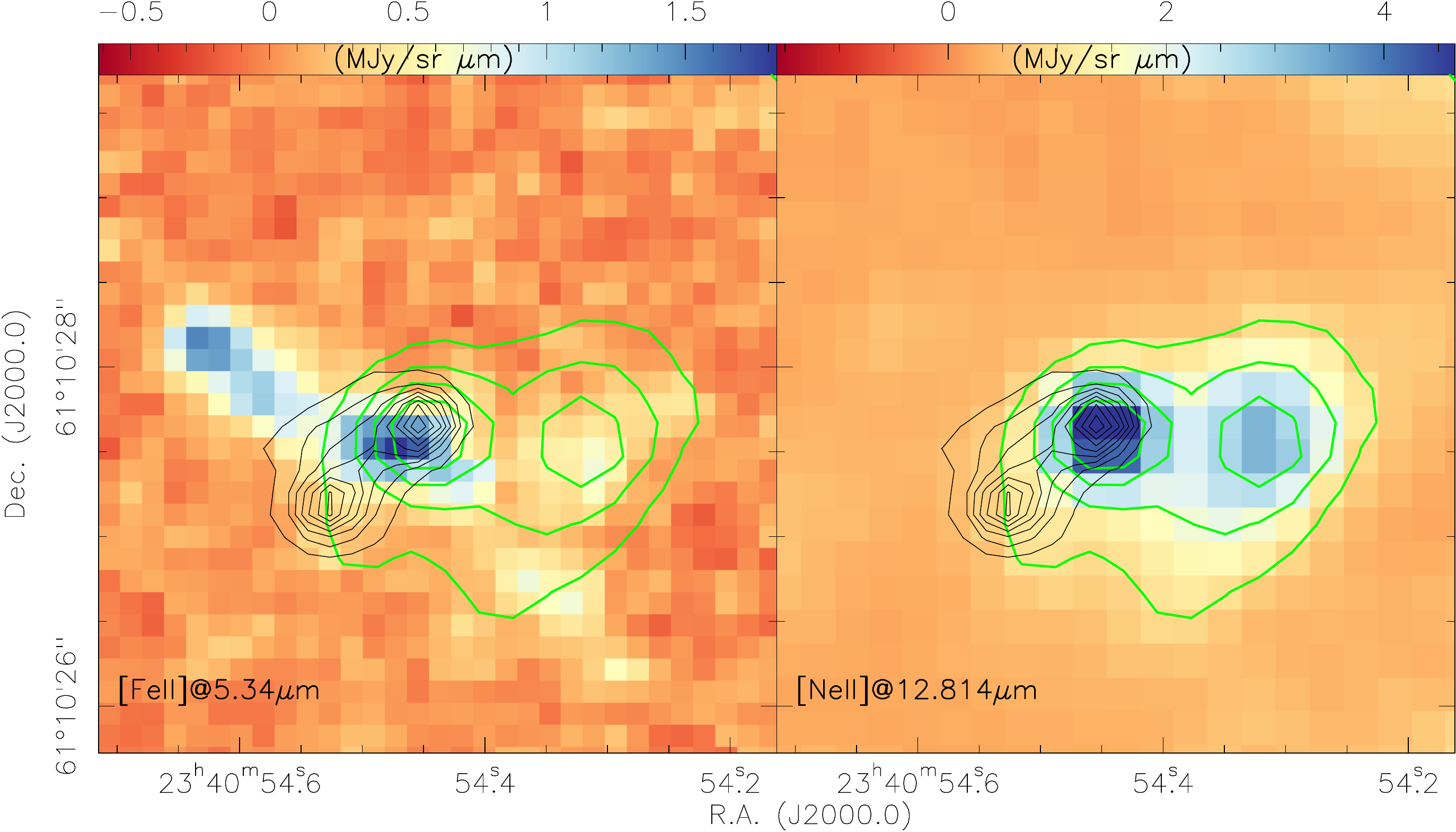}
\caption{Integrated [FeII]($^4F_{9/2}$--$^6D_{9/2}$) (left) and [NeII]($^2P_{1/2}-^2P_{3/2}$) (right) emission. The black contours show the 5.2\,$\mu$m continuum emission (steps 20 to 90\% by 10\%). The green contours outline again the [NeII] emission in $10\sigma$ steps ($1\sigma\sim 0.09$\,MJy\,sr$^{-1}\mu$m).} 
\label{neii} 
\end{figure*} 

Another potential jet-tracer is the [NeII]($^2P_{1/2}-^2P_{3/2}$) line at 12.814\,$\mu$m (e.g., \citealt{hollenbach1989,lefloch2003}). Figure \ref{neii} presents a comparison of the [NeII] and the [FeII] emission, and one finds two spatially very different distributions. While the [FeII] emission mainly traces the northeast-southwest oriented collimated jet-like structure, the [NeII] emission is rather double-peaked with one peak centered on mid-infrared source A, whereas the second [NeII] emission peak is offset by $\sim$0.9$''$ to the west. Comparing the [NeII] emission directly to the [FeII], one finds weak [FeII] emission associated also with the secondary [NeII] peak (left panel of Fig.~\ref{neii}). This [FeII] emission arises mainly at relatively high projected red-shifted velocities (see channels at $-$22 and +22\,km\,s$^{-1}$ in Fig.~\ref{channel_fe}). We note that no clear velocity structure can be spectrally resolved in the [NeII] data. The [NeII] emission can stem from EUV and/or X-ray-irradiated surface layers of accretion disks as well as from jets and outflows (e.g., \citealt{hollenbach1989,lefloch2003,glassgold2007,pascucci2009,guedel2010}). The J-shock models by \citet{hollenbach1989} predict that the [NeII] emission should strongly increase for gas velocities greater than $\sim$80\,km\,s$^{-1}$. In addition to this, high densities are also increasing the [NeII] intensities in these shock models (Fig.~7 in \citealt{hollenbach1989}). The spatial correspondence of the high-velocity red-shifted [FeII] and [NeII] emission close to the center of the region, where gas densities are also highest, is consistent with these J-shock models. In addition to this, these regions close to the central protostars, traced by the mid-infrared continuum emission, should also be exposed most to UV radiation. Hence a combination  of fast shocks and UV radiation may favor the [NeII] emission in that region. The non-detection of compact [NeII] emission in the rest of the IRAS\,23385+6053 region, where partly also high gas velocities and/or dense gas are observed (e.g., Figs.~\ref{channel_h2} and \ref{overview}), confirms that not only high velocities and dense gas are important, but that the missing UV radiation further away from the main protostars is likely to play an important role as well.

In summary, the combination of these datasets reveals at least three outflows, one in the northwest-southeast direction (I), another in the northeast-southwest direction (III) and at least one more in the north-south direction (IIa \& IIb). The fact that the H$_2$ emission shows two emission peaks in the north can be explained by either an expanding outflow cavity of one outflow or alternatively by two more jet-like structures. Differentiating between these two scenarios is not possible with the given data. The three, or potentially even four outflows are marked with white lines in Fig.~\ref{outflows} (right panel).

\subsection{Humphreys $\alpha$ line emission as an accretion tracer?}
\label{accretion_obs}

\begin{figure}[htb]
\includegraphics[width=0.49\textwidth]{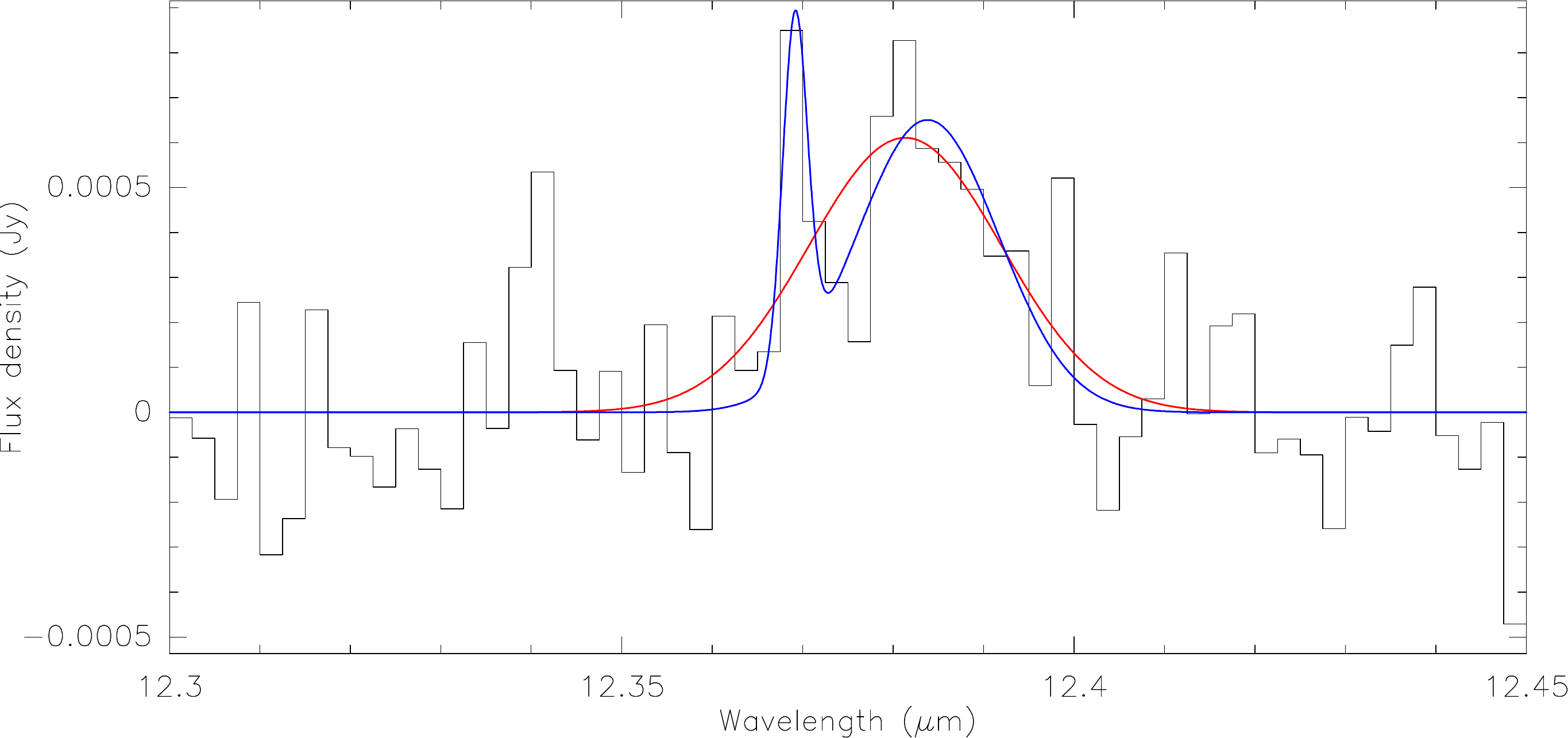}
\caption{H{\sc i}(7--6) spectrum at 12.37\,$\mu$m centered right in the middle between sources A and B (Fig.~\ref{continuum}), and averaged over an aperture with radius $2''$. The red line presents a single Gaussian fit to the spectrum whereas the blue line shows a two component Gaussian fit to the spectrum.}
\label{halpha} 
\end{figure}

One crucial missing parameter in high-mass star formation is the actual accretion rate onto the protostars. Here, we differentiate the accretion rate onto the protostar from otherwise more typically reported gas infall rates within the protostellar envelopes and cores (e.g, \citealt{myers1996,mardones1997,beuther2013b,wyrowski2016,vandishoeck2021}). While for low-mass T Tauri stars the accretion rate can be measured with optical or near-infrared lines (e.g., \citealt{connelley2010,salyk2013,hartmann2016,alcala2017}), for the deeply embedded phases in low- and high-mass star formation, one has to resort to longer wavelengths. Here, we focus on the Humphreys $\alpha$ line H{\sc{i}}(7--6) at 12.37\,$\mu$m.  \citet{rigliaco2015} explored this H{\sc i}(7--6) line at 12.37\,$\mu$m in Spitzer data of classical T Tauri stars, and they found a tentative correlation between the integrated H{\sc i}(7--6) luminosity and the accretion luminosity. We now explore this relation also for IRAS\,23385+6053. Figure \ref{halpha} presents the spectrum of the H{\sc i}(7--6) line extracted from the position inbetween mid-infrared sources A and B, and averaged over an aperture with radius of $2''$. While the red line shows a single Gaussian fit, the blue line presents a two component Gaussian fit. It should be noted that at this stage we consider this only a tentative detection, even though the peak flux density of the single-Gaussian fit with a peak flux density of 0.61\,mJy reaches a signal-to-noise ratio $S/N$ of 3 (in the channel with highest flux densities even $S/N\sim4.1$). The peak flux densities of the two-component fit (peak flux densities of 0.78 and 0.65\,mJy, respectively) have nominal $S/N$ ratios of 3.8 and 3.2, respectively.

With a nominal central wavelength for the H{\sc i}(7--6) line of 12.3719\,$\mu$m, the two components are shifted by $-70$ and $+293$\,km\,s$^{-1}$, respectively. The full-width-half-maximum line-widths of the two components are $\sim$73 and $\sim$436\,km\,s$^{-1}$, respectively. While the two-component nature of the spectrum may be attributed to gas associated with the innermost red- and blue-shifted rotating gas, the line-width of several hundred km\,s$^{-1}$ is consistent with gas at almost free-fall onto the central protostar. The free-fall velocity of gas around a 9\,$M_{\odot}$ protostar \citep{cesaroni2019} with a radius of 5\,$R_{\odot}$ would be $\sim$829\,km\,s$^{-1}$, but is likely narrower if the protostar were significantly bloated (e.g., \citealt{hosokawa2009}).

Assuming this to be a real detection, we can derive the integrated flux values of the one- and two-component fits. Since the integrated values of the different fits are similar, the following analysis is done for the two-component fit. However, with the single-component fit one gets comparable results. The integrated line luminosity of the two-component Gaussian fit is $\sim 2.3\times 10^{-4}$\,$L_{\odot}$. This value lies at the high-luminosity end of the relation fitted by \citet{rigliaco2015} to a sample of low-mass protostars. This then could be interpreted as an indicator that this relation may also extend to the high-mass regime. 

Following equation 1 in \citet{rigliaco2015}, one can convert the line luminosity to an accretion luminosity of 142\,$L_{\odot}$. Assuming now that all that luminosity is from infalling gas that converts its energy to radiation, one can infer an estimate of the accretion rate, e.g., Eq.~11.5 in \citet{stahler2005}. Taking these numbers at face-value, for an 9\,$M_{\odot}$ protostar \citep{cesaroni2019} with a radius of 5\,$R_{\odot}$ we can estimate an accretion rate of $\sim 2.6\times 10^{-6}$\,$M_{\odot}$yr$^{-1}$. 

The above fitted integrated line luminosity is a lower limit since extinction may weaken the real line emission.  We now estimate an extinction correction using the extinction curve derived by \citet{mcclure2009}. Based on the 9.7\,$\mu$m silicate absorption feature, Rocha et al.~(in prep.) estimate an extinction $A_V\sim 30-40$\,mag. Using $A_V\sim$30\,mag, and following \citet{mcclure2009}, that extinction can be converted to a K-band $A_K$ and then 12.37\,$\mu$m extinction values of 3.8 and 1.86\,mag, respectively. Such a 12.37\,$\mu$m extinction would boost the integrated H{\sc{i}}(7--6) line luminosity by a factor $\sim$5.56. Using the same approach as above, one can estimate now an extinction-corrected accretion luminosity of $\sim 5070$\,$L_{\odot}$, and an accretion rate of $\sim 0.9\times 10^{-4}$\,$M_{\odot}$yr$^{-1}$. The latter is roughly a factor $\sim$36 higher than the estimate without the extinction correction.

One caveat is that the extinction-corrected accretion luminosity of $\sim 5070$\,$L_{\odot}$ is on the same order as the luminosity estimated for the source of $\sim$3000\,$L_{\odot}$ \citep{molinari2008b}. Because of that, higher accretion luminosities would be unreasonable, and correspondingly, we do not consider higher extinction values. This accretion luminosity would imply that most of the bolometric luminosity originates from accretion, thus the source could be in its main accretion phase. If the approach outlined by \citet{rigliaco2015} and followed here is also applicable for young high-mass protostars, the extinction-corrected accretion rate estimate has to be an upper limit to the real accretion rate onto the protostar. The accretion rates will be discussed in depth in Sect.~\ref{accretion_discussion}.

\section{Discussion}

\subsection{Embedded protostars}
\label{protostars}

What is the nature of the two mid-infrared sources A and B? Are they individual protostars or are they potentially part of an extended disk-like structure perpendicular to the northeast-southwest outflow best depicted in the [FeII] data (Fig.~\ref{outflows})? 

The spatial separation between sources A and B of $0.67''$ corresponds at the measured distance of 4.9\,kpc to a linear separation of $\sim$3300\,au. This seems too large to be consistent with typical disk sizes, even around high-mass stars (e.g., \citealt{kuiper2010}).

Furthermore, we can inspect the full MIRI spectra towards the continuum sources. Fig.~\ref{spectrum} is extracted within a diameter of $2.4''$ combining sources A and B. At the shorter wavelengths, where the sources can still be spatially separated, the main difference between A and B are an even deeper silicate absorption feature towards source B which is also manifested in the non-detection of source B in the continuum images at 9.0 and 10.65\,$\mu$m (Fig.~\ref{continuum}).
Comparing the MIRI spectrum to typical spectra from embedded protostars to evolved disks (e.g., \citealt{evans2003,gibb2004c}), the MIRI spectrum presented here shows all the features typical for embedded protostars, from deep silicate absorption features around 10\,$\mu$m to the rising spectral energy distribution towards long wavelengths to several ice features at shorter wavelengths. Hence, the spectra support the assessment that sources A and B are two separate embedded protostars. 

As outlined in \citet{cesaroni2019}, the IRAS\,23385+6053 star-formation complex is a high-mass cluster in formation. They identify already six potential star-forming cores from the 1.3\,mm continuum emission (Fig.~5 in \citealt{cesaroni2019}, see also the right panel of our Fig.~\ref{overview}). The main mm peak position -- source mmA1 in \citet{cesaroni2019}, see Figs.~\ref{overview}, \ref{continuum} and \ref{outflows} -- has no obvious mid-infrared counterpart. However, the main mm peak is elongated in the southwestern direction (mm source mmA2 in \citealt{cesaroni2019}), and this mm-peak mmA2 is likely the counterpart to the mid-infrared source A identified with JWST. Ahmadi et al.~(subm.) analyze the CH$_3$CN mm line emission around the mm sources mmA1 and mmA2, and in their disk stability analysis they find very low Toomre Q values near mm peak mmA2 corresponding to the here identified mid-infrared source A. Ahmadi et al.~(subm.) argue that the low Toomre value could indicate a fragmenting disk-like structure. In that picture, mid-infrared source A could have potentially formed in that fragmented disk.

In comparison to that, the mid-infrared source B has no clear mm peak counterpart and hence can be considered as an independent entity, potentially of lower mass. Both the mm as well as the mid-infrared observations are sensitivity limited, hence more sources should likely exist below our detection limits that could be revealed by deep MIRI imaging.

Although the MIRI spectra are dominated by protostellar envelope features, the fact that we identify several outflows, and that two of them may likely emanate from the mid-infrared sources A and B (see the following Sect.~\ref{outflows_discussion}), the existence of embedded accretion disks is very likely. They are just difficult to identify because of the limited spatial resolution given the distance of the source (4.9\,kpc), and the spectra being dominated by the surrounding envelopes. In addition to that, rotational signatures indicative of an embedded disk around the main mm continuum source mmA1 was already derived from mm CH$_3$CN data \citep{cesaroni2019}.

\subsection{Outflows from mid-infrared to mm wavelengths}
\label{outflows_discussion}

The velocity ranges observed by the mid-infrared H$_2$ and [FeII] lines with more than $\pm 100$\,km\,s$^{-1}$ are much broader than those observed at mm wavelengths in the SiO emission of around $\pm 15$\,km\,s$^{-1}$ (Fig.~\ref{outflows}). Furthermore, the [NeII] emission requires high gas velocities as well (e.g., \citealt{hollenbach1989}). Therefore, the H$_2$, [FeII] and [NeII] lines are tracing genuinely higher velocity gas than the lines typically observed at mm wavelengths. This implies that the mid-infrared lines originate directly from the underlying jet, whereas the SiO emission rather traces outflowing gas at lower velocities. Since silicon (Si) has to be liberated from grains, some (slow) shocks are also needed for SiO emission (e.g., \citealt{schilke1997a,anderl2013}). Hence the SiO emission likely stems from a combination of low-velocity shocks caused by the jet, as well as entrained outflow gas. This behavior is also reflected by the excitation energies of the different lines. While the SiO(2--1) line has an upper level excitation energy $E_u/k$ of only 6.3\,K (e.g., \citealt{schoeier2005}), the presented H$_2$, [FeII] and [NeII] lines at 5.511, 5.34  and 12.814\,$\mu$m have much higher excitation energies $E_u/k$ $\sim$7202\,K, 2694\,K and 1123\,K, respectively\footnote{https://www.mpe.mpg.de/ir/ISO/linelists/Molecular.html, https://www.pa.uky.edu/$\sim$peter/newpage/}. Hence, the mid-infrared lines trace ionized as well as neutral shocked material, either from the jet itself or also from the outflow cavity walls. In contrast, the mm lines, even SiO that is typically also assumed to be a shock tracer (e.g., \citealt{schilke1997a}), traces colder components of the jet, where molecules can form at the internal working surfaces (e.g., \citealt{santiago-garcia2009}). The lesson is that studying the jets and outflows with a single tracer may miss some components. In contrast to that, the combined mid-infrared and mm data reveal a much more complete picture of the dynamical outflows driven from various embedded protostars. Furthermore, this outflow multiplicity also enforces the picture of high-mass stars forming exclusively in a clustered mode.

The more extended integral-shaped SiO emission in Fig.~\ref{outflows} corresponds to the northwest-southeast outflow (I). Could that be the signature of a precessing outflow or jet? Different jet-precession scenarios are discussed in, e.g., \citet{fendt1998}, and a potential precession reason is the outflow-driving within a binary or multiple system. Since we are dealing with a cluster-forming region and several identified protostars (Sect.~\ref{protostars}), precession induced by multiplicity may indeed explain the observed bent SiO emission.

With multiple outflows, it is also important to assess the potential driving sources for each of them. Inspecting Figure \ref{outflows} again, the [FeII] jet (III) is most closely centered on the mid-infrared source A, but neither directly spatially aligned with mid-infrared source B or mm peak mmA1. Therefore, the most likely driving source for outflow (III) appears to be the mid-infrared source A (spatially coincident with the secondary mm-peak mmA2). 

Regarding the northwest-southeastern outflow (I), \citet{cesaroni2019} proposed a rather straight orientation center on a mm source C (identified in the H$_2$CO emission, Fig.~\ref{overview} right panel) at the southwestern edge of the mm continuum emission (in the middle between the red and blue outflow arrows in Fig.~\ref{overview}, right panel). Considering the more bent structure we infer from the SiO and H$_2$ emission (Fig.~\ref{outflows}), the driving source is more likely close to the mid-infrared sources A and B. Since source A is the likely driver of outflow (III), we propose mid-infrared source B as a candidate driving source for outflow (I).  

Finally, the one (or potentially two) north-south outflows (IIa \& IIb) are spatially associated with mid-infrared sources A and B as well as the main mm-peak mmA1. Since mid-infrared sources A and B are more likely linked to outflows (I) and (III), we propose the mm-peak mmA1 to host the driving source(s) of these north-south outflow(s).

\subsection{Accretion rates}
\label{accretion_discussion}

How do the Humphreys $\alpha$ H{\sc{i}}(7--6) inferred accretion rate estimates between $2.6\times 10^{-6}$ and $0.9\times 10^{-4}$\,$M_{\odot}$yr$^{-1}$ (Sect.~\ref{accretion_obs}) compare to outflow rates? \citet{molinari1998b} report integrated outflow rates for IRAS\,23385+6053 of $\geq 10^{-3}$\,$M_{\odot}$yr$^{-1}$ based on HCO$^+$ data. Based on their lower spatial resolution, these outflow rates likely combine the different outflows we see in the new data. Assuming momentum conservation between initial jet and entrained outflow rate, and a velocity ratio between jet and entrained gas of $\sim$20 (e.g., \citealt{beuther2002b}), the jet-mass-flow rate is roughly an order of magnitude lower. Furthermore, jet models result in ratios between jet-flow rate and accretion rate of roughly 0.3 (e.g., \citealt{tomisaka1998,shu1999}). Hence, the measured outflow rates should correspond to accretion rates on the order 10$^{-4}$\,$M_{\odot}$yr$^{-1}$. While such accretion rate estimates are almost two orders of magnitude higher than the non-extinction-corrected H{\sc{i}}(7--6) accretion rate estimate, they are in the same ballpark as those derived when considering the extinction correction.

\citet{cesaroni2019} also estimated accretion rates based on the assumption that all cores are collapsing in free-fall, and they get accretion rates between a few times $10^{-4}$ and a few times $10^{-3}$\,$M_{\odot}$yr$^{-1}$. The corresponding time-scales they estimate from the accretion rates are mostly below $10^4$\,yrs. While these accretion rate estimates are again at the upper end of our H{\sc{i}}(7--6) based estimates, \citet{cesaroni2019} point out that their estimated time-scales are too short for typical high-mass star formation times on the order $10^5$\,yr. Hence, their estimated time-scales are lower limits, and correspondingly their estimated accretion rates upper limits.

Our H{\sc{i}}(7--6) estimated accretion rates appear at the lower end of high-mass star formation accretion rates if one considers that accretion rates on the order $10^{-4}-10^{-3}$\,$M_{\odot}$yr$^{-1}$ are needed to form a high-mass star within a few hundred thousand years (e.g., \citealt{mckee2003}). While we cannot entirely exclude that the main object is not a 9\,$M_{\odot}$ protostar but maybe a multiple system containing lower-mass objects, this seems  unlikely since the luminosity of $3\times 10^3$\,$L_{\odot}$ \citep{molinari2008b} requires a more massive central object (e.g., \citealt{mottram2011,cesaroni2019}). The luminosity is also consistent with the estimated $\sim$9\,$M_{\odot}$ embedded object based on the rotation curves measured in CH$_3$CN \citep{cesaroni2019}. In comparison to that, in the extinction-corrected accretion estimate, almost the entire luminosity should stem from the accretion processes. 

Furthermore, with a total mass reservoir of the IRAS\,23385+6053 star-forming region of $\sim$510\,$M_{\odot}$, and assuming a typical initial mass function and star formation efficiency, no star above 10\,$M_{\odot}$ would be expected in this region (e.g., \citealt{tackenberg2012}). 
One can estimate, for example, the maximum stellar mass $m^*_{\rm{max}}$ of a cluster following \citet{larson2003}, \citet{sanhueza2019} and Morii et al.~(subm.):

$$ m^*_{\rm{max}} = 15.6 \left(\frac{M_{\rm{clump}}}{10^3\rm{M}_{\odot}} \frac{\epsilon_{\rm{SFE}}}{0.3}\right)^{0.45} M_{\odot}.$$

With a clump mass $M_{\rm{clump}}=510$\,$M_{\odot}$ and a star formation efficiency $\epsilon_{\rm{SFE}}\sim 0.15$, the resulting maximum stellar mass $m^*_{\rm{max}}$ would be $\sim$8.45\,$M_{\odot}$, roughly what is found for IRAS23385+6053 \citep{cesaroni2019}. Therefore, it could well be that the main accretion phase is coming to an end and hence we see comparably lower accretion rates in the H{\sc{i}}(7--6) line. A similar decrease of accretion rates with time is also seen in low-mass regions (e.g., \citealt{evans2009}).

While the outflow rate, and by that the inferred accretion rate, is an integral measure over the outflow time-scale, the H{\sc{i}}(7--6) inferred accretion rate should rather correspond to an instantaneous accretion rate onto the star at the time of observations. Considering that a 9\,$M_{\odot}$ protostar has already formed at the center of IRAS\,23385+6053, it may well be that the outflow-inferred accretion rates correspond to the active accretion phase that comes to an end now. In that picture, the H{\sc{i}}(7--6) accretion rate gives an estimate of the still ongoing remaining accretion. In addition to this, (high-mass) star formation is known to be episodic (e.g., \citealt{caratti2017,kueffmeier2017,hunter2021,elbakyan2021}), and in that framework IRAS\,23395+6053 may also be in a low-accretion phase at the time of observations.

Some potential caveats need to be discussed. The original relation between accretion luminosity and H{\sc{i}}(7--6) line luminosity was inferred for low-mass protostars, and it is not obvious that this relation has to hold also for the high-mass regime. The H{\sc{i}}(7--6) line is a recombination line, and while that should arise in the low-mass regime mainly from accretion processes, high-mass protostars can also emit UV radiation that could result in recombination line emission. However, towards IRAS\,23385+6053, only extended cm continuum emission, most likely associated with the extended mid-infrared nebula (Fig.~\ref{overview}), has been detected \citep{molinari2002}, but no compact cm free-free emission towards the central protostar could be identified \citep{molinari1998b}. Therefore, no strong ionizing radiation is emitted from the embedded high-mass protostar. Furthermore, in that evolutionary stage, high-mass protostars can also be bloated and then have much lower surface temperatures (e.g., \citealt{hosokawa2009,hosokawa2010}), again reducing the potential ionizing radiation. Therefore, from a conceptional point of view, in the early evolutionary stages of high-mass star formation where the protostar can be bloated and does not yet emit much ionizing radiation, the relation between accretion luminosity and H{\sc{i}}(7--6) line luminosity may still be valid.

An additional caveat is that the H{\sc{i}}(7--6) detection is only at a $\sim 3-4\sigma$ level. Other HI emission lines (e.g., H{\sc{i}}(6--5) or H{\sc{i}}(8--7)) are also in the observed bandpass but remain undetected in IRAS23395+6053. Therefore, more observations of other high-mass star-forming regions are required to explore the capability of tracing accretion with the H{\sc{i}}(7--6) and other recombination lines in more depth. This will become possible, given the fact that most JOYS targets are still pending in the JWST queue. With a sample of approximately two dozen regions, sufficient data will become available to derive statistically unambiguous values. The present work is a first step in realizing this. 

\section{Conclusions and summary}

We present among the first JWST MIRI MRS mid-infrared observations of a young high-mass star-forming region, IRAS\,23385+6053. The spectral coverage between $\sim$5 and $\sim$28\,$\mu$m reveals a plethora of spectral gas lines from atomic and molecular species. Furthermore, several broader ice-features can be identified that will be studied in accompanying papers. 


Investigating the mid-infrared continuum emission, two mid-infrared sources are identified at the shortest wavelength with a separation of $\sim$0.67$''$ (or $\sim$3280\,au). At the long-wavelength end of the spectrum, these two sources merge into one emission structure because of the lower angular resolution. While one of the mid-infrared sources (A) is associated with a mm continuum source (mmA2 in \citealt{cesaroni2019}), the other mid-infrared source B has no obvious mm counterpart. The MIRI spectra of the two sources clearly show their deeply embedded protostellar nature. Combining the JWST mid-infrared and previously obtained mm data confirms that we are indeed observing a cluster in its making.

For the outflow analysis, we focus on two shorter wavelength lines to take advantage of the high spatial resolution. Combining the outflow-tracing JWST MIRI H$_2$ at 5.511\,$\mu$m and [FeII] at 5.34\,$\mu$m lines with the mm SiO(2--1) emission, we identify at least three independent outflows in the region, and it is possible to assign individual candidate driving sources to each of them. Furthermore, although only observed at a spectral resolving power of $\sim$3500, corresponding to a velocity-resolution of $\sim$86\,km\,s$^{-1}$, with given velocity shifts of $\pm 140$\,km\,s$^{-1}$, we can spatially and spectrally resolve the red- and blue-shifted high-velocity gas. In particular the red-shifted high-velocity gas is spatially associated with red-shifted SiO emission, although the latter is at lower velocities of $\sim$15\,km\,s$^{-1}$. Furthermore, the potential additional mid-infrared outflow tracer [NeII] shows a partially overlapping morphology, with two emission peaks close to the center of the region. The two [NeII] emission structures can be associated with high velocity gas also seen in the [FeII] emission from the northwest-southeast outflow (I), but do not fully reproduce all observed [FeII] structures. These data show that the higher excited mid-infrared data trace also the higher-velocity jet-like structures whereas the lower-excited mm lines preferentially trace lower-velocity jet and entrained outflow gas of the broader molecular outflow.

Our investigation of a potential accretion rate tracer, the Humphreys $\alpha$ H{\sc{i}}(7--6) line, revealed a 3--4$\sigma$ detection. Using the relation between integrated H{\sc{i}}(7--6) line luminosity and accretion luminosity, and depending on extinction-correction, we can estimate accretion luminosities between 142 and 5070\,$L_{\odot}$. Assuming furthermore a reasonable protostellar mass and radius, these luminosities convert to accretion rate estimates between $\sim$2.6$\times 10^{-6}$ and $\sim$0.9$\times 10^{-4}$\,$M_{\odot}$yr$^{-1}$. Setting this into context with an accretion rate estimates based on outflow observations, IRAS\,23385+6053 may already be close to the end of the main accretion phase. However, with data for one region so far and then only a 3--4$\sigma$ detection, these results have to be taken with a bit of caution. Observations of more high-mass star-forming regions are needed to further evaluate the possibility of estimating accretion rates with the H{\sc{i}}(7--6) line. Nevertheless, the present work shows that the Humphreys $\alpha$ line has the potential to become an important accretion rate tracer with JWST. Future MIRI observations, also within the JOYS program, will set much tighter constraints on that.

In summary, these first JWST MIRI MRS observations of a high-mass star-forming region reveal important new results of the protostellar distribution, the outflow properties as well as accretion rates in IRAS\,23385+6053. The complementary atomic and molecular gas lines as well as the ice features are discussed in companion papers. While interesting in themselves, these data also show the enormous potential for JWST to boost star formation research.

\begin{acknowledgements}
 The following National and International Funding Agencies funded and
supported the MIRI development: NASA; ESA; Belgian Science Policy
Office (BELSPO); Centre Nationale d’Etudes Spatiales (CNES); Danish
National Space Centre; Deutsches Zentrum fur Luftund Raumfahrt (DLR);
Enterprise Ireland; Ministerio De Economi\'a y Competividad;
Netherlands Research School for Astronomy (NOVA); Netherlands
Organisation for Scientific Research (NWO); Science and Technology
Facilities Council; Swiss Space Office; Swedish National Space Agency;
and UK Space Agency.  We thank Fabiana Faustini for providing the infrared data. We also thank Roy van Boekel for mid-infrared line diagnostic discussions. H.B.~acknowledges support from the Deutsche Forschungsgemeinschaft in the Collaborative Research Center (SFB 881) “The Milky Way System” (subproject B1). EvD, MvG, LF, KS, WR and HL acknowledge support from ERC Advanced grant 101019751 MOLDISK, TOP-1 grant 614.001.751 from the Dutch Research Council (NWO), the Netherlands Research School for Astronomy (NOVA), the Danish National Research Foundation through the Center of Excellence “InterCat” (DNRF150), and DFG-grant 325594231, FOR 2634/2. P.J.K.~acknowledges financial support from the Science Foundation Ireland/Irish Research Council Pathway programme under Grant Number 21/PATH-S/9360. A.C.G. has been supported by PRIN-INAF MAIN-STREAM 2017 “Protoplanetary disks seen through the eyes of new- genera-tion instruments” and from PRIN-INAF 2019 “Spectroscopically tracing the disk dispersal evolution (STRADE)”. K.J.~acknowledges the support from the Swedish National Space Agency (SNSA). T.H.~acknowledges support from the European Research Council under the Horizon 2020 Framework Program via the ERC Advanced Grant "Origins" 83 24 28.
     
\end{acknowledgements}


\begin{thebibliography}{64}
\expandafter\ifx\csname natexlab\endcsname\relax\def\natexlab#1{#1}\fi

\bibitem[{{Alcal{\'a}} {et~al.}(2017){Alcal{\'a}}, {Manara}, {Natta}, {Frasca},
  {Testi}, {Nisini}, {Stelzer}, {Williams}, {Antoniucci}, {Biazzo}, {Covino},
  {Esposito}, {Getman}, \& {Rigliaco}}]{alcala2017}
{Alcal{\'a}}, J.~M., {Manara}, C.~F., {Natta}, A., {et~al.} 2017, \aap, 600,
  A20

\bibitem[{{An} {et~al.}(2011){An}, {Ram{\'{\i}}rez}, {Sellgren}, {Arendt},
  {Adwin Boogert}, {Robitaille}, {Schultheis}, {Cotera}, {Smith}, \&
  {Stolovy}}]{an2011}
{An}, D., {Ram{\'{\i}}rez}, S.~V., {Sellgren}, K., {et~al.} 2011, \apj, 736,
  133

\bibitem[{{Anderl} {et~al.}(2013){Anderl}, {Guillet}, {Pineau des For{\^e}ts},
  \& {Flower}}]{anderl2013}
{Anderl}, S., {Guillet}, V., {Pineau des For{\^e}ts}, G., \& {Flower}, D.~R.
  2013, \aap, 556, A69

\bibitem[{{Beuther} {et~al.}(2013){Beuther}, {Linz}, \&
  {Henning}}]{beuther2013b}
{Beuther}, H., {Linz}, H., \& {Henning}, T. 2013, \aap, 558, A81

\bibitem[{{Beuther} {et~al.}(2018){Beuther}, {Mottram}, {Ahmadi}, {Bosco},
  {Linz}, {Henning}, {Klaassen}, {Winters}, {Maud}, {Kuiper}, {Semenov},
  {Gieser}, {Peters}, {Urquhart}, {Pudritz}, {Ragan}, {Feng}, {Keto},
  {Leurini}, {Cesaroni}, {Beltran}, {Palau}, {S{\'a}nchez-Monge},
  {Galvan-Madrid}, {Zhang}, {Schilke}, {Wyrowski}, {Johnston}, {Longmore},
  {Lumsden}, {Hoare}, {Menten}, \& {Csengeri}}]{beuther2018b}
{Beuther}, H., {Mottram}, J.~C., {Ahmadi}, A., {et~al.} 2018, \aap, 617, A100

\bibitem[{{Beuther} {et~al.}(2002){Beuther}, {Schilke}, {Sridharan}, {Menten},
  {Walmsley}, \& {Wyrowski}}]{beuther2002b}
{Beuther}, H., {Schilke}, P., {Sridharan}, T.~K., {et~al.} 2002, \aap, 383, 892

\bibitem[{Bushouse {et~al.}(2022)Bushouse, Eisenhamer, Dencheva, Davies,
  Greenfield, Morrison, Hodge, Simon, Grumm, Droettboom, Slavich, Sosey, Pauly,
  Miller, Jedrzejewski, Hack, Davis, Crawford, Law, Gordon, Regan, Cara,
  MacDonald, Bradley, Shanahan, Jamieson, Teodoro, \& Williams}]{bushouse2022}
Bushouse, H., Eisenhamer, J., Dencheva, N., {et~al.} 2022, {JWST Calibration
  Pipeline}

\bibitem[{{Caratti o Garatti} {et~al.}(2017){Caratti o Garatti}, {Stecklum},
  {Garcia Lopez}, {Eisl{\"o}ffel}, {Ray}, {Sanna}, {Cesaroni}, {Walmsley},
  {Oudmaijer}, {de Wit}, {Moscadelli}, {Greiner}, {Krabbe}, {Fischer}, {Klein},
  \& {Iba{\~n}ez}}]{caratti2017}
{Caratti o Garatti}, A., {Stecklum}, B., {Garcia Lopez}, R., {et~al.} 2017,
  Nature Physics, 13, 276

\bibitem[{{Casoli} {et~al.}(1986){Casoli}, {Dupraz}, {Gerin}, {Combes}, \&
  {Boulanger}}]{casoli1986}
{Casoli}, F., {Dupraz}, C., {Gerin}, M., {Combes}, F., \& {Boulanger}, F. 1986,
  \aap, 169, 281

\bibitem[{{Cesaroni} {et~al.}(2019){Cesaroni}, {Beuther}, {Ahmadi},
  {Beltr{\'a}n}, {Csengeri}, {Galv{\'a}n-Madrid}, {Gieser}, {Henning},
  {Johnston}, {Klaassen}, {Kuiper}, {Leurini}, {Linz}, {Longmore}, {Lumsden},
  {Maud}, {Moscadelli}, {Mottram}, {Palau}, {Peters}, {Pudritz},
  {S{\'a}nchez-Monge}, {Schilke}, {Semenov}, {Suri}, {Urquhart}, {Winters},
  {Zhang}, \& {Zinnecker}}]{cesaroni2019}
{Cesaroni}, R., {Beuther}, H., {Ahmadi}, A., {et~al.} 2019, \aap, 627, A68

\bibitem[{{Cesaroni} {et~al.}(1988){Cesaroni}, {Palagi}, {Felli}, {Catarzi},
  {Comoretto}, {Di Franco}, {Giovanardi}, \& {Palla}}]{cesaroni1988}
{Cesaroni}, R., {Palagi}, F., {Felli}, M., {et~al.} 1988, \aaps, 76, 445

\bibitem[{{Connelley} \& {Greene}(2010)}]{connelley2010}
{Connelley}, M.~S. \& {Greene}, T.~P. 2010, \aj, 140, 1214

\bibitem[{{Elbakyan} {et~al.}(2021){Elbakyan}, {Nayakshin}, {Vorobyov},
  {Caratti o Garatti}, \& {Eisl{\"o}ffel}}]{elbakyan2021}
{Elbakyan}, V.~G., {Nayakshin}, S., {Vorobyov}, E.~I., {Caratti o Garatti}, A.,
  \& {Eisl{\"o}ffel}, J. 2021, \aap, 651, L3

\bibitem[{{Evans} {et~al.}(2003){Evans}, {Allen}, {Blake}, {Boogert}, {Bourke},
  {Harvey}, {Kessler}, {Koerner}, {Lee}, {Mundy}, {Myers}, {Padgett},
  {Pontoppidan}, {Sargent}, {Stapelfeldt}, {van Dishoeck}, {Young}, \&
  {Young}}]{evans2003}
{Evans}, Neal~J., I., {Allen}, L.~E., {Blake}, G.~A., {et~al.} 2003, \pasp,
  115, 965

\bibitem[{{Evans} {et~al.}(2009){Evans}, {Dunham}, {J{\o}rgensen}, {Enoch},
  {Mer{\'{\i}}n}, {van Dishoeck}, {Alcal{\'a}}, {Myers}, {Stapelfeldt},
  {Huard}, {Allen}, {Harvey}, {van Kempen}, {Blake}, {Koerner}, {Mundy},
  {Padgett}, \& {Sargent}}]{evans2009}
{Evans}, N.~J., {Dunham}, M.~M., {J{\o}rgensen}, J.~K., {et~al.} 2009, \apjs,
  181, 321

\bibitem[{{Faustini} {et~al.}(2009){Faustini}, {Molinari}, {Testi}, \&
  {Brand}}]{faustini2009}
{Faustini}, F., {Molinari}, S., {Testi}, L., \& {Brand}, J. 2009, \aap, 503,
  801

\bibitem[{{Fendt} \& {Zinnecker}(1998)}]{fendt1998}
{Fendt}, C. \& {Zinnecker}, H. 1998, \aap, 334, 750

\bibitem[{{Gibb} {et~al.}(2004){Gibb}, {Whittet}, {Boogert}, \&
  {Tielens}}]{gibb2004c}
{Gibb}, E.~L., {Whittet}, D.~C.~B., {Boogert}, A.~C.~A., \& {Tielens},
  A.~G.~G.~M. 2004, \apjs, 151, 35

\bibitem[{{Gibb} {et~al.}(2000){Gibb}, {Whittet}, {Schutte}, {Boogert},
  {Chiar}, {Ehrenfreund}, {Gerakines}, {Keane}, {Tielens}, {van Dishoeck}, \&
  {Kerkhof}}]{gibb2000}
{Gibb}, E.~L., {Whittet}, D.~C.~B., {Schutte}, W.~A., {et~al.} 2000, \apj, 536,
  347

\bibitem[{{Gieser} {et~al.}(2021){Gieser}, {Beuther}, {Semenov}, {Ahmadi},
  {Suri}, {M{\"o}ller}, {Beltr{\'a}n}, {Klaassen}, {Zhang}, {Urquhart},
  {Henning}, {Feng}, {Galv{\'a}n-Madrid}, {de Souza Magalh{\~a}es},
  {Moscadelli}, {Longmore}, {Leurini}, {Kuiper}, {Peters}, {Menten},
  {Csengeri}, {Fuller}, {Wyrowski}, {Lumsden}, {S{\'a}nchez-Monge}, {Maud},
  {Linz}, {Palau}, {Schilke}, {Pety}, {Pudritz}, {Winters}, \&
  {Pi{\'e}tu}}]{gieser2021}
{Gieser}, C., {Beuther}, H., {Semenov}, D., {et~al.} 2021, \aap, 648, A66

\bibitem[{{Glassgold} {et~al.}(2007){Glassgold}, {Najita}, \&
  {Igea}}]{glassgold2007}
{Glassgold}, A.~E., {Najita}, J.~R., \& {Igea}, J. 2007, \apj, 656, 515

\bibitem[{{Guedel} {et~al.}(2010){Guedel}, {Lahuis}, {Briggs}, {Carr},
  {Glassgold}, {Henning}, {Najita}, {van Boekel}, \& {van
  Dishoeck}}]{guedel2010}
{Guedel}, M., {Lahuis}, F., {Briggs}, K.~R., {et~al.} 2010, ArXiv e-prints
  [\eprint[arXiv]{1006.2848}]

\bibitem[{{Hartmann} {et~al.}(2016){Hartmann}, {Herczeg}, \&
  {Calvet}}]{hartmann2016}
{Hartmann}, L., {Herczeg}, G., \& {Calvet}, N. 2016, \araa, 54, 135

\bibitem[{{Hollenbach} \& {McKee}(1989)}]{hollenbach1989}
{Hollenbach}, D. \& {McKee}, C.~F. 1989, \apj, 342, 306

\bibitem[{{Hosokawa} \& {Omukai}(2009)}]{hosokawa2009}
{Hosokawa}, T. \& {Omukai}, K. 2009, \apj, 691, 823

\bibitem[{{Hosokawa} {et~al.}(2010){Hosokawa}, {Yorke}, \&
  {Omukai}}]{hosokawa2010}
{Hosokawa}, T., {Yorke}, H.~W., \& {Omukai}, K. 2010, \apj, 721, 478

\bibitem[{{Hunter} {et~al.}(2021){Hunter}, {Brogan}, {De Buizer}, {Towner},
  {Dowell}, {MacLeod}, {Stecklum}, {Cyganowski}, {El-Abd}, \&
  {McGuire}}]{hunter2021}
{Hunter}, T.~R., {Brogan}, C.~L., {De Buizer}, J.~M., {et~al.} 2021, \apjl,
  912, L17

\bibitem[{{Kuffmeier} {et~al.}(2017){Kuffmeier}, {Haugb{\o}lle}, \&
  {Nordlund}}]{kueffmeier2017}
{Kuffmeier}, M., {Haugb{\o}lle}, T., \& {Nordlund}, {\r{A}}. 2017, \apj, 846, 7

\bibitem[{{Kuiper} {et~al.}(2010){Kuiper}, {Klahr}, {Beuther}, \&
  {Henning}}]{kuiper2010}
{Kuiper}, R., {Klahr}, H., {Beuther}, H., \& {Henning}, T. 2010, \apj, 722,
  1556

\bibitem[{{Kurtz} {et~al.}(2004){Kurtz}, {Hofner}, \&
  {{\'A}lvarez}}]{kurtz2004}
{Kurtz}, S., {Hofner}, P., \& {{\'A}lvarez}, C.~V. 2004, \apjs, 155, 149

\bibitem[{{Labiano} {et~al.}(2021){Labiano}, {Argyriou},
  {{\'A}lvarez-M{\'a}rquez}, {Glasse}, {Glauser}, {Patapis}, {Law}, {Brandl},
  {Justtanont}, {Lahuis}, {Mart{\'\i}nez-Galarza}, {Mueller}, {Noriega-Crespo},
  {Royer}, {Shaughnessy}, \& {Vandenbussche}}]{labiano2021}
{Labiano}, A., {Argyriou}, I., {{\'A}lvarez-M{\'a}rquez}, J., {et~al.} 2021,
  \aap, 656, A57

\bibitem[{{Larson}(2003)}]{larson2003}
{Larson}, R.~B. 2003, Reports on Progress in Physics, 66, 1651

\bibitem[{{Lefloch} {et~al.}(2003){Lefloch}, {Cernicharo}, {Cabrit},
  {Noriega-Crespo}, {Moro-Mart{\'\i}n}, \& {Cesarsky}}]{lefloch2003}
{Lefloch}, B., {Cernicharo}, J., {Cabrit}, S., {et~al.} 2003, \apjl, 590, L41

\bibitem[{{Mardones} {et~al.}(1997){Mardones}, {Myers}, {Tafalla}, {Wilner},
  {Bachiller}, \& {Garay}}]{mardones1997}
{Mardones}, D., {Myers}, P.~C., {Tafalla}, M., {et~al.} 1997, \apj, 489, 719

\bibitem[{{McClure}(2009)}]{mcclure2009}
{McClure}, M. 2009, \apjl, 693, L81

\bibitem[{{McClure} {et~al.}(2023){McClure}, {Rocha}, {Pontoppidan}, {Crouzet},
  {Chu}, {Dartois}, {Lamberts}, {Noble}, {Pendleton}, {Perotti}, {Qasim},
  {Rachid}, {Smith}, {Sun}, {Beck}, {Boogert}, {Brown}, {Caselli}, {Charnley},
  {Cuppen}, {Dickinson}, {Drozdovskaya}, {Egami}, {Erkal}, {Fraser}, {Garrod},
  {Harsono}, {Ioppolo}, {Jimenez-Serra}, {Jin}, {J{\o}rgensen}, {Kristensen},
  {Lis}, {McCoustra}, {McGuire}, {Melnick}, {Oberg}, {Palumbo}, {Shimonishi},
  {Sturm}, {van Dishoeck}, \& {Linnartz}}]{mcclure2023}
{McClure}, M.~K., {Rocha}, W.~R.~M., {Pontoppidan}, K.~M., {et~al.} 2023, arXiv
  e-prints, arXiv:2301.09140

\bibitem[{{McKee} \& {Tan}(2003)}]{mckee2003}
{McKee}, C.~F. \& {Tan}, J.~C. 2003, \apj, 585, 850

\bibitem[{{Molinari} {et~al.}(2008){Molinari}, {Faustini}, {Testi}, {Pezzuto},
  {Cesaroni}, \& {Brand}}]{molinari2008b}
{Molinari}, S., {Faustini}, F., {Testi}, L., {et~al.} 2008, \aap, 487, 1119

\bibitem[{{Molinari} {et~al.}(1998){Molinari}, {Testi}, {Brand}, {Cesaroni}, \&
  {Palla}}]{molinari1998b}
{Molinari}, S., {Testi}, L., {Brand}, J., {Cesaroni}, R., \& {Palla}, F. 1998,
  \apjl, 505, L39

\bibitem[{{Molinari} {et~al.}(2002){Molinari}, {Testi}, {Rodr{\'{\i}}guez}, \&
  {Zhang}}]{molinari2002}
{Molinari}, S., {Testi}, L., {Rodr{\'{\i}}guez}, L.~F., \& {Zhang}, Q. 2002,
  \apj, 570, 758

\bibitem[{{Motte} {et~al.}(2018){Motte}, {Bontemps}, \& {Louvet}}]{motte2018}
{Motte}, F., {Bontemps}, S., \& {Louvet}, F. 2018, \araa, 56, 41

\bibitem[{{Mottram} {et~al.}(2011){Mottram}, {Hoare}, {Urquhart}, {Lumsden},
  {Oudmaijer}, {Robitaille}, {Moore}, {Davies}, \& {Stead}}]{mottram2011}
{Mottram}, J.~C., {Hoare}, M.~G., {Urquhart}, J.~S., {et~al.} 2011, \aap, 525,
  A149

\bibitem[{{Myers} {et~al.}(1996){Myers}, {Mardones}, {Tafalla}, {Williams}, \&
  {Wilner}}]{myers1996}
{Myers}, P.~C., {Mardones}, D., {Tafalla}, M., {Williams}, J.~P., \& {Wilner},
  D.~J. 1996, \apjl, 465, L133

\bibitem[{{Palla} {et~al.}(1991){Palla}, {Brand}, {Cesaroni}, {Comoretto}, \&
  {Felli}}]{palla1991}
{Palla}, F., {Brand}, J., {Cesaroni}, R., {Comoretto}, G., \& {Felli}, M. 1991,
  \aap, 246, 249

\bibitem[{{Pascucci} \& {Sterzik}(2009)}]{pascucci2009}
{Pascucci}, I. \& {Sterzik}, M. 2009, \apj, 702, 724

\bibitem[{{Pontoppidan} {et~al.}(2022){Pontoppidan}, {Barrientes}, {Blome},
  {Braun}, {Brown}, {Carruthers}, {Coe}, {DePasquale}, {Espinoza}, {Marin},
  {Gordon}, {Henry}, {Hustak}, {James}, {Jenkins}, {Koekemoer}, {LaMassa},
  {Law}, {Lockwood}, {Moro-Martin}, {Mullally}, {Pagan}, {Player}, {Proffitt},
  {Pulliam}, {Ramsay}, {Ravindranath}, {Reid}, {Robberto}, {Sabbi}, {Ubeda},
  {Balogh}, {Flanagan}, {Gardner}, {Hasan}, {Meinke}, \&
  {Nota}}]{pontoppidan2022}
{Pontoppidan}, K.~M., {Barrientes}, J., {Blome}, C., {et~al.} 2022, \apjl, 936,
  L14

\bibitem[{{Rigliaco} {et~al.}(2015){Rigliaco}, {Pascucci}, {Duchene},
  {Edwards}, {Ardila}, {Grady}, {Mendigut{\'\i}a}, {Montesinos}, {Mulders},
  {Najita}, {Carpenter}, {Furlan}, {Gorti}, {Meijerink}, \&
  {Meyer}}]{rigliaco2015}
{Rigliaco}, E., {Pascucci}, I., {Duchene}, G., {et~al.} 2015, \apj, 801, 31

\bibitem[{{Salyk} {et~al.}(2013){Salyk}, {Herczeg}, {Brown}, {Blake},
  {Pontoppidan}, \& {van Dishoeck}}]{salyk2013}
{Salyk}, C., {Herczeg}, G.~J., {Brown}, J.~M., {et~al.} 2013, \apj, 769, 21

\bibitem[{{Sanhueza} {et~al.}(2019){Sanhueza}, {Contreras}, {Wu}, {Jackson},
  {Guzm{\'a}n}, {Zhang}, {Li}, {Lu}, {Silva}, {Izumi}, {Liu}, {Miura},
  {Tatematsu}, {Sakai}, {Beuther}, {Garay}, {Ohashi}, {Saito}, {Nakamura},
  {Saigo}, {Veena}, {Nguyen-Luong}, \& {Tafoya}}]{sanhueza2019}
{Sanhueza}, P., {Contreras}, Y., {Wu}, B., {et~al.} 2019, \apj, 886, 102

\bibitem[{{Santiago-Garc{\'\i}a} {et~al.}(2009){Santiago-Garc{\'\i}a},
  {Tafalla}, {Johnstone}, \& {Bachiller}}]{santiago-garcia2009}
{Santiago-Garc{\'\i}a}, J., {Tafalla}, M., {Johnstone}, D., \& {Bachiller}, R.
  2009, \aap, 495, 169

\bibitem[{{Schilke} {et~al.}(1997){Schilke}, {Walmsley}, {Pineau des Forets},
  \& {Flower}}]{schilke1997a}
{Schilke}, P., {Walmsley}, C.~M., {Pineau des Forets}, G., \& {Flower}, D.~R.
  1997, \aap, 321, 293

\bibitem[{{Sch{\"o}ier} {et~al.}(2005){Sch{\"o}ier}, {van der Tak}, {van
  Dishoeck}, \& {Black}}]{schoeier2005}
{Sch{\"o}ier}, F.~L., {van der Tak}, F.~F.~S., {van Dishoeck}, E.~F., \&
  {Black}, J.~H. 2005, \aap, 432, 369

\bibitem[{{Shu} {et~al.}(1999){Shu}, {Allen}, {Shang}, {Ostriker}, \&
  {Li}}]{shu1999}
{Shu}, F.~H., {Allen}, A., {Shang}, H., {Ostriker}, E.~C., \& {Li}, Z. 1999, in
  NATO ASIC Proc. 540: The Origin of Stars and Planetary Systems, 193

\bibitem[{{Stahler} \& {Palla}(2005)}]{stahler2005}
{Stahler}, S.~W. \& {Palla}, F. 2005, {The Formation of Stars} (ISBN
  3-527-40559-3.~Wiley-VCH)

\bibitem[{{Tackenberg} {et~al.}(2012){Tackenberg}, {Beuther}, {Henning},
  {Schuller}, {Wienen}, {Motte}, {Wyrowski}, {Bontemps}, {Bronfman}, {Menten},
  {Testi}, \& {Lefloch}}]{tackenberg2012}
{Tackenberg}, J., {Beuther}, H., {Henning}, T., {et~al.} 2012, \aap, 540, A113

\bibitem[{{Tomisaka}(1998)}]{tomisaka1998}
{Tomisaka}, K. 1998, \apjl, 502, L163

\bibitem[{{van Dishoeck} {et~al.}(2021){van Dishoeck}, {Kristensen}, {Mottram},
  {Benz}, {Bergin}, {Caselli}, {Herpin}, {Hogerheijde}, {Johnstone}, {Liseau},
  {Nisini}, {Tafalla}, {van der Tak}, {Wyrowski}, {Baudry}, {Benedettini},
  {Bjerkeli}, {Blake}, {Braine}, {Bruderer}, {Cabrit}, {Cernicharo}, {Choi},
  {Coutens}, {de Graauw}, {Dominik}, {Fedele}, {Fich}, {Fuente}, {Furuya},
  {Goicoechea}, {Harsono}, {Helmich}, {Herczeg}, {Jacq}, {Karska}, {Kaufman},
  {Keto}, {Lamberts}, {Larsson}, {Leurini}, {Lis}, {Melnick}, {Neufeld},
  {Pagani}, {Persson}, {Shipman}, {Taquet}, {van Kempen}, {Walsh}, {Wampfler},
  {Y{\i}ld{\i}z}, \& {WISH Team}}]{vandishoeck2021}
{van Dishoeck}, E.~F., {Kristensen}, L.~E., {Mottram}, J.~C., {et~al.} 2021,
  \aap, 648, A24

\bibitem[{{van Dishoeck} {et~al.}(1998){van Dishoeck}, {Wright}, {Cernicharo},
  {Gonz{\'a}lez-Alfonso}, {de Graauw}, {Helmich}, \&
  {Vandenbussche}}]{vandishoeck1998b}
{van Dishoeck}, E.~F., {Wright}, C.~M., {Cernicharo}, J., {et~al.} 1998, \apjl,
  502, L173

\bibitem[{{Whittet} {et~al.}(1996){Whittet}, {Schutte}, {Tielens}, {Boogert},
  {de Graauw}, {Ehrenfreund}, {Gerakines}, {Helmich}, {Prusti}, \& {van
  Dishoeck}}]{whittet1996}
{Whittet}, D.~C.~B., {Schutte}, W.~A., {Tielens}, A.~G.~G.~M., {et~al.} 1996,
  \aap, 315, L357

\bibitem[{{Wolf-Chase} {et~al.}(2012){Wolf-Chase}, {Smutko}, {Sherman},
  {Harper}, \& {Medford}}]{wolf-chase2012}
{Wolf-Chase}, G., {Smutko}, M., {Sherman}, R., {Harper}, D.~A., \& {Medford},
  M. 2012, \apj, 745, 116

\bibitem[{{Wouterloot} \& {Brand}(1989)}]{wouterloot1989}
{Wouterloot}, J.~G.~A. \& {Brand}, J. 1989, \aaps, 80, 149

\bibitem[{{Wright} {et~al.}(2010){Wright}, {Eisenhardt}, {Mainzer}, {Ressler},
  {Cutri}, {Jarrett}, {Kirkpatrick}, {Padgett}, {McMillan}, {Skrutskie},
  {Stanford}, {Cohen}, {Walker}, {Mather}, {Leisawitz}, {Gautier}, {McLean},
  {Benford}, {Lonsdale}, {Blain}, {Mendez}, {Irace}, {Duval}, {Liu}, {Royer},
  {Heinrichsen}, {Howard}, {Shannon}, {Kendall}, {Walsh}, {Larsen}, {Cardon},
  {Schick}, {Schwalm}, {Abid}, {Fabinsky}, {Naes}, \& {Tsai}}]{wright2010}
{Wright}, E.~L., {Eisenhardt}, P.~R.~M., {Mainzer}, A.~K., {et~al.} 2010, \aj,
  140, 1868

\bibitem[{{Wyrowski} {et~al.}(2016){Wyrowski}, {G{\"u}sten}, {Menten},
  {Wiesemeyer}, {Csengeri}, {Heyminck}, {Klein}, {K{\"o}nig}, \&
  {Urquhart}}]{wyrowski2016}
{Wyrowski}, F., {G{\"u}sten}, R., {Menten}, K.~M., {et~al.} 2016, \aap, 585,
  A149

\bibitem[{{Yang} {et~al.}(2022){Yang}, {Green}, {Pontoppidan}, {Bergner},
  {Cleeves}, {Evans}, {Garrod}, {Jin}, {Kim}, {Kim}, {Lee}, {Sakai},
  {Shingledecker}, {Shope}, {Tobin}, \& {van Dishoeck}}]{yang2022}
{Yang}, Y.-L., {Green}, J.~D., {Pontoppidan}, K.~M., {et~al.} 2022, \apjl, 941,
  L13

\end{thebibliography}



\end{document}